\documentclass[10pt, a4paper]{article}
\usepackage[round]{natbib}
\usepackage[hidelinks]{hyperref}
\usepackage{booktabs} 
\usepackage[font=small,labelfont=bf]{caption} 
\usepackage{amsfonts, amsmath, amsthm, amssymb}
\usepackage{indentfirst} 
\usepackage{algorithmic}
\usepackage{ascmac}
\usepackage{url}
\usepackage{threeparttable}

\usepackage{graphicx}
\usepackage{caption}
\usepackage{subcaption}

\begin{document}
\title{\textbf{Identification in Bayesian Estimation of the Skewness Matrix in a Multivariate Skew-Elliptical Distribution}\thanks{This research is supported by the Keio University Doctorate Student Grant-in-Aid Program from Ushioda Memorial Fund; and JSPS KAKENHI under Grant [number MKK337J].}}\author{Sakae Oya\footnote{\url{sakae.prosperity21@keio.jp}}\\Graduate School of Economics, Keio University, Tokyo, Japan \\and\\ Teruo Nakatsuma\\Faculty of Economics, Keio University, Tokyo, Japan}
\date{}
\maketitle
\begin{abstract}
Harvey et al. (2010) extended the Bayesian estimation method by Sahu et al. (2003) to a multivariate skew-elliptical distribution with a general skewness matrix, and applied it to Bayesian portfolio optimization with higher moments. Although their method is epochal in the sense that it can handle the skewness dependency among asset returns and incorporate higher moments into portfolio optimization, it cannot identify all elements in the skewness matrix due to label switching in the Gibbs sampler. To deal with this identification issue, we propose to modify their sampling algorithm by imposing a positive lower-triangular constraint on the skewness matrix of the multivariate skew-elliptical distribution and improved interpretability. Furthermore, we propose a Bayesian sparse estimation of the skewness matrix with the horseshoe prior to further improve the accuracy. In the simulation study, we demonstrate that the proposed method with the identification constraint can successfully estimate the true structure of the skewness dependency while the existing method suffers from the identification issue.

\end{abstract}

\section{Introduction}
\label{intro}
The mean-variance approach proposed by \cite{Markowitz} still plays the central role in portfolio management even today. One of the key assumptions of this approach is that asset returns jointly follow a multivariate normal distribution, though it is well-known that they tend to follow a fat-tailed, possibly skewed distribution as \cite{Kon}, \cite{Mills}, \cite{Markowitz96}, \cite{Peiro} among others have pointed out. Therefore, researches have proposed numerous distributions that can express these characteristics of asset returns well. In particular, a so-called skew-t distribution is often assumed for asset returns since Hansen (1994) first used it for modeling financial data. There are various types of skew-t distribution known in the literature, but arguably the most famous one is based on the generalized hyperbolic (GH) distribution.

The GH distribution, which was originally introduced by \cite{BN77}, can flexibly describe many distributions including the normal distribution, hyperbolic distribution, normal inverse Gaussian (NIG) distribution, Student's t distribution, and skew-t distribution. The skew-t distribution as a special case of the GH distribution is called the GH skew-t distribution. \cite{Hansen}, \cite{Fernandez} and \cite{AasHaff} assumed the GH skew-t distribution for asset returns. Especially, application of the GH distribution has been recently advanced in the field of asset price volatility models. For example, \cite{NakajimaOmori} assumed the GH skew-t distribution for the error distribution of the stochastic volatility (SV) model and proposed a Bayesian Markov chain Monte Carlo (MCMC) method while \cite{Nakajima2017} constructed a sparse estimation method for the skewness parameter of the GH skew-t distribution in the SV model and demonstrated that it could improve prediction accuracy.

Although the GH distribution is flexible enough to model a single asset on many occasions, it has difficulty in capturing the skewness dependency among multiple assets. Fund managers would find the skewness dependency useful in particular when the financial market crashes and almost all assets suddenly go south since such sharp price co-movement may not be captured by the second moment (i.e., correlation) only.

To circumvent this shortcoming of the GH distribution, we propose to use the skew-elliptical distribution, which was proposed by \cite{BrancoDey} as a generalization of the multivariate skew-normal distribution by \cite{Azzalini96} and later improved by \cite{Sahu03}\footnote{Although we take up the skew-elliptical distribution based on \cite{Sahu03} with application to portfolio management in mind, there are alternative skew-elliptical-type distributions known in the literature. Research on skew-elliptical-type distributions to financial data is very active (e.g., \cite{BarbiRomagnoli}, \cite{CarmichaelCoen}, \cite{AR2014}). \cite{AdcockAzzalini2020} reviews the recent development in this field and explains relationship among various types of skew-elliptical distribution in detail.}. The skew-elliptical distribution includes the normal distribution, Student's t distribution and their skewed counterparts: skew-normal and skew-t distribution. Unlike the GH distribution, it is straightforward to extend the skew-elliptical distribution to the multivariate case. The multivariate skew-normal distribution has another advantage: Its Bayesian estimation can be conducted via pure Gibbs sampling. For example, \cite{Sahu03} proposed a Gibbs sampler for a linear regression model in which the error term follows a skew-elliptical distribution without skewness dependency. Moreover, \cite{Harvey2010} improved \cite{Sahu03}'s method, and applied it to Bayesian estimation of the multivariate skew-normal distribution as well as portfolio optimization that considers up to the third moment in the presence of skewness dependency.

In our assessment, however, the Bayesian estimation method of the multivariate skew-elliptical distribution by \cite{Harvey2010} has an identification issue about the skewness parameters due to so-called label switching. To elaborate on our point, let us look into the definition of a multivariate skew-elliptical distribution. For simplicity, we only consider the multivariate skew-normal distribution\footnote{In essence, any skew-elliptical distributions have the same identification issue. So we start with the skew-normal distribution as a representative example. It is straightforward to extend our argument to any skew-elliptical distributions including the skew-t distribution which we will deal with in Section 3.}. Suppose a $N\times 1$ random vector $R_t$ $(t=1,\dots,T)$ of asset returns follows a multivariate skew-normal distribution such that
\begin{align}
\label{sn.def1}
R_t &= \mu + \Delta Z_t + \epsilon_t,\\
Z_t & \sim \mathcal{N}^+(0,I_N),\quad \epsilon_t\sim \mathcal{N}(0,\Omega^{-1}),\nonumber \\ 
Z_t& \perp \epsilon_t, \nonumber
\end{align}
where each element in $Z_t$ is supposed to independently follow a positive half normal distribution with the scale parameter equal to 1. $\Delta$ and $\Omega$ are the skewness matrix\footnote{Here we call $\Delta$ the skewness matrix though it does not match the skewness of the distribution in this model. For more information, see Section 2.2 and Appendix A of \cite{Harvey2010}.} and the precision matrix\footnote{\cite{Harvey2010} used the covariance matrix in their specification of the skew-normal distribution and assumed the inverse-Wishart prior for it, which is equivalent to assuming the Wishart prior for the precision matrix in our specification. We use the precision matrix because we later examine the extended model that incorporates sparsity into the graphical structure among asset returns.} respectively. \cite{Harvey2010} did not impose any restriction and assumed $\Delta$ is full matrix
:
\[
    \Delta = \begin{bmatrix}
        \delta_{11} & \delta_{12}  &  \delta_{13}  &  \cdots & \delta_{1N} \\
        \delta_{21} & \delta_{22} &  \delta_{23}  &  \cdots  &  \delta_{2N} \\
        \delta_{31} & \delta_{32} & \delta_{33} &   \cdots   &  \delta_{3N}   \\
        \vdots      & \vdots      & \vdots      &     &  \vdots           \\
        \delta_{N1} & \delta_{N2} & \delta_{N3} & \cdots & \delta_{NN} \\
    \end{bmatrix}.
\]
By defining
\[
    R = \begin{bmatrix}
        R_1^{\intercal} \\ \vdots \\ R_T^{\intercal}
    \end{bmatrix},\quad
    \tilde R  = \begin{bmatrix}
        \tilde R_1^{\intercal} \\ \vdots \\ \tilde R_T^{\intercal}
    \end{bmatrix}
    = \begin{bmatrix}
        (R_1-\mu)^{\intercal} \\ \vdots \\ (R_T-\mu)^{\intercal}
    \end{bmatrix},\quad
    Z = \begin{bmatrix}
        Z_1^{\intercal} \\ \vdots \\ Z_T^{\intercal}
    \end{bmatrix},\quad
    E = \begin{bmatrix}
        \epsilon_1^{\intercal} \\ \vdots \\ \epsilon_T^{\intercal}
    \end{bmatrix},
\]
\eqref{sn.def1} can be rewritten as
\begin{align}
    \label{sn.def2}
    \tilde R & = Z \Delta^{\intercal} + E
\end{align}
Note that $Z\Delta^{\intercal}$ in \eqref{sn.def2} is
\begin{align}
\label{ZDelta}
    Z \Delta^{\intercal}& = \begin{bmatrix}
        Z_{11} & Z_{12}  &  Z_{13}  &  \cdots & Z_{1N} \\
        Z_{21} & Z_{22} &  Z_{23}  &  \cdots  &  Z_{2N} \\
        Z_{31} & Z_{32} & Z_{33} &   \cdots   &  Z_{3N}   \\
        \vdots      & \vdots      & \vdots      &     &  \vdots           \\
        Z_{T1} & Z_{T2} & Z_{T3} & \cdots & Z_{TN} \\
    \end{bmatrix}      
    \begin{bmatrix}
        \delta_{11} & \delta_{21}  &  \delta_{31}  &  \cdots & \delta_{N1} \\
        \delta_{12} & \delta_{22} &  \delta_{32}  &  \cdots  &  \delta_{N2} \\
        \delta_{13} & \delta_{23} & \delta_{33} &   \cdots   &  \delta_{N3}   \\
        \vdots      & \vdots      & \vdots      &     &  \vdots           \\
        \delta_{1N} & \delta_{2N} & \delta_{3N} & \cdots & \delta_{NN} \\
    \end{bmatrix}\\ \nonumber
     &=    \begin{bmatrix}
        Z_{11}\delta_{11} + Z_{12}\delta_{12}+ \cdots + Z_{1N}\delta_{1N} &    \cdots &   Z_{11}\delta_{N1} + Z_{12}\delta_{N2}+ \cdots + Z_{1N}\delta_{NN} \\
         Z_{21}\delta_{11} + Z_{22}\delta_{12}+ \cdots + Z_{2N}\delta_{1N} &    \cdots &   Z_{21}\delta_{N1} + Z_{22}\delta_{N2}+ \cdots + Z_{2N}\delta_{NN} \\
          \vdots      &    &  \vdots           \\
          Z_{T1}\delta_{11} + Z_{T2}\delta_{12}+ \cdots + Z_{TN}\delta_{1N} &    \cdots &   Z_{T1}\delta_{N1} + Z_{T2}\delta_{N2}+ \cdots + Z_{TN}\delta_{NN} \\  
       \end{bmatrix}.
\end{align}
Since the summation in each element of \eqref{ZDelta} is invariant in terms of permutation, the likelihood of $\Delta$ in the model \eqref{sn.def2} takes the same value for any permutations of the columns in $\Delta$. As a result, it is likely that the columns of $\Delta$ are randomly misaligned during the Gibbs sampler and their interpretability is lost. This problem is well-known in the field of latent factor models, which have a structure similar to the model \eqref{sn.def2}.

As far as we know, no research\footnote{\cite{PS2010} pointed out that, in the model by \cite{Sahu03} or \cite{Azzalini2003}, it becomes difficult to identify the parameter when the skewness parameter approaches to 0, and proposed the improved model with sparsity. The identification issue we point out in this paper still occurs regardless of the magnitude of the skewness parameter when the co-skewness is taken into consideration as in \cite{Harvey2010}. Note that this is a separate issue from \cite{PS2010}. In this paper as well, we will study an extended model with sparsity of co-skewness in Section 4.} has examined the identification issue of \cite{Harvey2010}'s model due to the label switching problem yet. Therefore we aim to construct a modified model in which the identification issue of $\Delta$ is resolved and the interpretability is assured. Moreover, we also propose an extended model assuming a shrinkage prior to further to improve the estimation accuracy.

This paper is organized as follows. In Section 2, we briefly review the estimation method by \cite{Harvey2010} and propose the modified method that solves the identification issue. Then we extend our proposed method by applying the shrinkage prior to the co-skewness. In Section 3, we perform simulation studies in multiple settings of the structure of $\Delta$ and verify whether proposed methods can properly estimate the true structure. The conclusion is given in Section 4.

\section{Proposed Method}
\label{model0}
First we review the Bayesian MCMC method proposed by \cite{Harvey2010}. Based on \eqref{sn.def1} and \eqref{sn.def2}, two equivalent expressions of the joint conditional density of $R$ given $Z$ is obtained as:
\begin{align}
    p(R|\mu,\Delta,\Omega,Z)
    \label{sn.likelihood1}
    &\propto |\Omega|^{\frac{T}2}\exp\left[-\frac12\sum_{t=1}^T(R_t-\mu-\Delta Z_t)^{\intercal}\Omega(R_t-\mu-\Delta Z_t)\right] \\
    \label{sn.likelihood2}
    &\propto |\Omega|^{\frac{T}2}\exp\left[-\frac12 tr \left\{\Omega(\tilde R-Z\Delta^{\intercal})^{\intercal}(\tilde R-Z\Delta^{\intercal})\right\}\right].
\end{align}
\cite{Harvey2010} assumed the following normal-Wishart prior for $\mu$, $\delta$ and $\Omega$\footnote{While \cite{Harvey2010} sampled $\mu$ and $\Delta$ together by jointly assuming the multivariate normal prior for them, we describe $\mu$ and $\Delta$ separately because we will later extend our proposed method to the model with a shrinkage prior for $\Delta$.}:
\begin{equation}
    \label{prior0}
    \mu \sim \mathcal{N}(b_\mu,A_\mu^{-1}),\quad
    \Delta \sim \mathcal{N}(b_\Delta, A_\Delta^{-1}),\quad \Omega \sim \mathcal{W}(S_\Omega^{-1},\nu_\Omega).
\end{equation}
We refer to the skew elliptical distribution with the normal- Wishart prior \eqref{prior0} as Full-NOWI. With Bayes' theorem, the posterior distribution of $(\mu,\Delta,\Omega)$ is obtained as
\begin{equation}
    \label{posterior0}
    p(\mu,\Delta,\Omega|R) \propto \int_0^\infty\cdots\int_0^\infty p(R|\mu,\Delta,\Omega,Z)p(Z_1)dZ_1\cdots p(Z_T)dZ_T p(\mu)p(\Delta)p(\Omega).
\end{equation}
Since the multiple integral in \eqref{posterior0} is intractable, we employ Monte Carlo integration to compute the summary statistics of parameters in the posterior distribution \eqref{posterior0}. For this purpose, we apply a Markov chain sampling method to draw the latent variables $(Z_1,\dots,Z_T)$ along with the parameters $(\mu,\Delta,\Omega)$ from the posterior distribution \eqref{posterior0}.

The full conditional posterior distribution of $\mu$, $\Delta$, $\Omega$, and $Z_{t}$ are derived as follows.
\begin{align}
    \label{fc.mu}
    \mu|\cdot &\sim \mathcal{N}\left(\hat A_\mu^{-1}\hat b_\mu,\hat A_\mu^{-1}\right),\quad
    \hat A_\mu = A_\mu + T\Omega,\quad
    \hat b_\mu = A_\mu b_\mu + \Omega(R-Z\Delta^{\intercal})^{\intercal}\iota, \\
    \label{fc.Delta}
    \Delta|\cdot &\sim \mathcal{N}(\hat A_\Delta^{-1}\hat b_\Delta,\hat A_\Delta^{-1}),\quad \hat A_\Delta = A_\Delta + Z^{\intercal}\tilde\Omega Z,\quad \hat b_\Delta = A_\Delta b_\Delta + Z^{\intercal}\tilde\Omega y, \\
    \label{fc.omega}
    \Omega|\cdot &\sim \mathcal{W}\left(\hat S^{-1}, \hat\nu\right),\quad \hat\nu = \nu_\Omega + T,\quad \hat S = S_\Omega + S,\quad S = (\tilde R-Z\Delta^{\intercal})^{\intercal}(\tilde R-Z\Delta^{\intercal}), \\
    \label{fc.latent}
    Z_t|\cdot &\sim \mathcal{N}^+\left(\hat A_z^{-1}\hat b_z, \hat A_z^{-1}\right),\quad \hat A_z = I_N + \Delta^{\intercal}\Omega\Delta,\quad \hat b_z = \Delta^{\intercal}\Omega(R_t-\mu),
\end{align}
where $Z^{\intercal}\tilde\Omega Z = \sum_{t=1}^T Z_t^{\intercal}\Omega Z_t$ and $Z^{\intercal}\tilde\Omega y = \sum_{t=1}^T Z_t^{\intercal}\Omega\tilde R_t$.

Since it is difficult to jointly draw $Z_t$ from \eqref{fc.latent}, the element-wise Gibbs sampler can be applied to \eqref{fc.latent}. Without loss of generality, we partition $Z_t$, $\mu_z=\hat A_z^{-1}\hat b_z$ and $\hat A_z$ as
\[
    Z_t = \begin{bmatrix}
        z_{1t} \\ Z_{2t}
    \end{bmatrix},\quad
    \mu_z = \begin{bmatrix}
        \mu_{z1} \\ \mu_{z2}
    \end{bmatrix},\quad
    \hat A_z = \begin{bmatrix}
        a_{11} & a_{21}^{\intercal} \\
        a_{21} & A_{22}
    \end{bmatrix},
\]
where $z_{1t}$, $\mu_{z1}$ and $a_{11}$ are scalars, $Z_{2t}$, $\mu_{z2}$ and $a_{21}$ are $(N-1)\times 1$ vectors, and $A_{22}$ is an $(N-1)\times(N-1)$ matrix. Then the full conditional posterior distribution of $z_{1t}$ is
\begin{equation}
    \label{fc.latent.element}
    z_{1t}|\cdot \sim
    \mathcal{N}^+\left(\mu_{z1}-\frac1{a_{11}}a_{21}^{\intercal}(Z_{2t}-\mu_{z2}), \frac1{a_{11}}\right).
\end{equation}
The full conditional posterior distribution of the second to the last element of $Z_t$ can be derived in the same manner as \eqref{fc.latent.element}. Then we can construct the element-wise Gibbs sampler for $Z_t$ by drawing each element of $Z_t$ sequentially from its full conditional  posterior distribution.

Since columns in $\Delta$ are not identified without imposing any constraints as we confirmed in the introduction, we use  a positive lower-triangular constraint (PLT, \cite{GewekeZhou}, \cite{West2003} and \cite{LopesWest2004})\footnote{Although, \cite{SchnatterLopes2018} recently proposed a generalized lower triangular condition that generalizes the positive lower-triangular (GLT) condition, but in the case of the multivariate skew-elliptical distribution, the GLT condition matches the PLT condition since $\Delta$ is square matrix. Therefore, we used the PLT condition in this research.} on $\Delta$ which is often used in econometric field.  Assume upper-triangular above the main diagonal of $\Delta$ equals to zero as:
\begin{equation}
    \label{lt}
    \Delta = \begin{bmatrix}
        \delta_{11} &             &             &        &             \\
        \delta_{21} & \delta_{22} &             &        &             \\
        \delta_{31} & \delta_{32} & \delta_{33} &        &             \\
        \vdots      & \vdots      & \vdots      & \ddots &             \\
        \delta_{N1} & \delta_{N2} & \delta_{N3} & \cdots & \delta_{NN} \\
    \end{bmatrix}.
\end{equation}
By defining
\[
    \Delta Z_t = W_t\delta,\quad
    W_t = \begin{bmatrix}
        z_{1t} &        &        &        &        &        &        \\
               & z_{1t} & z_{2t} &        &        &        &        \\
               &        &        & \ddots &        &        &        \\
               &        &        &        & z_{1t} & \cdots & z_{Nt} \\
    \end{bmatrix},\quad
    \delta_t = \begin{bmatrix}
        \delta_{11} \\ \delta_{21} \\ \delta_{22} \\ \vdots \\ \delta_{N1} \\ \vdots \\ \delta_{NN}
    \end{bmatrix},
\]
we can rewrite \eqref{sn.def2} as 
\begin{equation}
    \label{sn.def3}
    y = W\delta + \epsilon,\quad \epsilon\sim\mathcal{N}(0,\tilde\Omega^{-1}),
\end{equation}
where 
\[
    y = vec(\tilde R^{\intercal}),\quad
    W = \begin{bmatrix} W_1 \\ \vdots \\ W_T \end{bmatrix},\quad
    \epsilon = vec(E^{\intercal}),\quad
    \tilde\Omega = I_T\otimes \Omega.
\]
Using W and $\delta$, \eqref{prior0} can be rewritten as:
\begin{equation}
    \label{prior}
    \mu \sim \mathcal{N}(b_\mu,A_\mu^{-1}),\quad
    \delta \sim \mathcal{N}(b_\delta, A_\delta^{-1}),\quad \Omega \sim \mathcal{W}(S_\Omega^{-1},\nu_\Omega).
\end{equation}
We refer to the multivariate skew-elliptical distribution with the lower-triangle constraint \eqref{lt} and the normal-Wishart prior \eqref{prior} as LT-NOWI.

With \eqref{sn.def3} and \eqref{prior}, the full conditional posterior distribution of $\delta$ is derived as:
\begin{equation}
    \label{fc.delta}
    \delta|\cdot \sim \mathcal{N}(\hat A_\delta^{-1}\hat b_\delta,\hat A_\delta^{-1}),\quad \hat A_\delta = A_\delta + W^{\intercal}\tilde\Omega W,\quad \hat b_\delta = A_\delta b_\delta + W^{\intercal}\tilde\Omega y,
\end{equation}
where $W^{\intercal}\tilde\Omega W = \sum_{t=1}^T W_t^{\intercal}\Omega W_t$ and $W^{\intercal}\tilde\Omega y = \sum_{t=1}^T W_t^{\intercal}\Omega\tilde R_t$. The posterior distribution of $\Omega$, $\mu$, $Z$ are the same as in \eqref{fc.omega}, \eqref{fc.mu} and \eqref{fc.latent}.

It is known that, when the normal-Wishart prior is used, the posterior distribution may not have a sharp peak around zero even if the true value is exactly equal to zero. In order to make the posterior distribution shrink toward zero and improve the estimation accuracy, we propose an extended model with a shrinkage prior for $\Delta$ and $\Omega$.

To non-zero elements in $\Delta$, we apply the horseshoe prior (\cite{Carvalho2010}):
\begin{equation}
    \label{delta.horseshoe1}
    \delta_j \sim \mathcal{N}(0,\lambda_j^2\tau^2),\quad \lambda_j \sim C^+(0,1),\quad \tau \sim C^+(0,1),\quad \left(j = 1,\dots, \frac{N(N+1)}{2}\right),
\end{equation}
where $\delta_j$ is the $j$-th element in $\delta$ and $C^+(\cdot)$ stands for the half Cauchy distribution. Note that the half Cauchy distribution in \eqref{delta.horseshoe1} is expressed as a mixture of inverse gamma distributions as in \cite{MakalicSchmidt2016}:
\begin{equation}
    \label{delta.horseshoe2}
     \lambda_j^2|\nu_j \sim IG\left(\frac12,\frac1{\nu_j}\right),\quad \tau^2|\xi \sim IG\left(\frac12,\frac1{\xi}\right),\quad \nu_j,\xi \sim IG\left(\frac12,1\right).
\end{equation}
For computation, we first randomly generate $\nu_j$, $\xi$ from the inverse Gaussian distribution. Then, we also randomly generate $\lambda_j^2$, $\tau^2$  and set them as initial values.
The derivation of the full conditional posterior distribution of $\delta$ is straightforward. Given $\lambda_1,\dots,\lambda_N,\tau$, the prior distribution of $\delta$ is
\[
    \delta|\lambda_1,\dots,\lambda_{N^2},\tau \sim \mathcal{N}\left(0,\tau^2\mathrm{diag}\left(\lambda_1^2,\dots,\lambda_{N^2}^2\right)\right).
\]
Thus, the full conditional posterior distribution of $\delta$ is identical to \eqref{fc.delta} except
\[
    A_\delta = \frac1{\tau^2}\mathrm{diag}\left(\frac1{\lambda_1^2},\dots,\frac1{\lambda_{N^2}^2}\right),\quad
    b_\delta = 0.
\]
The full conditional posterior distributions of $\lambda_j^2$ and $\tau^2$ in \eqref{delta.horseshoe1} are
\begin{align}
    \lambda_j^2|\cdot &\sim IG\left(1,\frac1{\nu_j}+\frac{\delta_j^2}{2\tau^2}\right),\quad (j=1,\dots,N^2), \\
        \tau^2|\cdot &\sim IG\left(\frac{N^2+1}2,\frac1{\xi}+\frac12\sum_{j=1}^{N^2}\frac{\delta_j^2}{\lambda_j^2}\right),
\end{align}
while those of the auxiliary variables are
\begin{align}
    \nu_j|\cdot &\sim IG\left(1,1+\frac1{\lambda_j^2}\right),\quad (j=1,\dots,N^2), \\
    \xi|\cdot &\sim IG\left(1,1+\frac1{\tau^2}\right).
\end{align}

We also apply the graphical horseshoe prior to the off-diagonal elements in $\Omega$ as in \cite{Li2019}. Although it is tempting to use a horseshoe prior such as
\begin{align}
    \label{omega.horseshoe1}
    \omega_{ij} &\sim \mathcal{N}(0,\rho_{ij}^2\psi^2),\quad (1 \leqq i < j \leqq N), \\
    \label{omega.horseshoe2}
    \rho_{ij}^2|\upsilon_{ij} &\sim IG\left(\frac12,\frac1{\upsilon_{ij}}\right),\quad \psi^2|\zeta \sim IG\left(\frac12,\frac1{\zeta}\right),\quad \upsilon_{ij}, \zeta \sim IG\left(\frac12,1\right),
\end{align}
where $\omega_{ij}$ is the $(i,j)$ element in $\Omega$, \eqref{omega.horseshoe1} is not appropriate for our purpose because the support of $(\omega_{ij})_{i<j}$ in \eqref{omega.horseshoe1} includes points where $\Omega$ is not positive definite. Thus we need to put the positive definiteness constraint upon \eqref{omega.horseshoe1}. In this paper, we refer to the multivariate skew-elliptical distribution with the lower-triangle constraint \eqref{lt}, the horseshoe prior for the skewness matrix $\Delta$ \eqref{delta.horseshoe1} and the positive-definiteness-assured graphical horseshoe prior for the precision matrix $\Omega$ \eqref{omega.horseshoe1}--\eqref{omega.horseshoe2} as LT-HSGHS.

To assure the positive definiteness of $\Omega$ in the course of sampling, we apply a block Gibbs sampler by \cite{ONGLASSO}. To illustrate the block Gibbs sampler, we introduce the following partition of $\Omega$ and $S$:
\begin{equation}
    \label{partition}
    \Omega = \begin{bmatrix}
        \omega_{11} & \omega_{21}^{\intercal} \\
        \omega_{21} & \Omega_{22}
    \end{bmatrix},\quad
    S = \begin{bmatrix}
        s_{11} & s_{21}^{\intercal} \\
        s_{21} & S_{22}
    \end{bmatrix},
\end{equation}
where $\omega_{11}$ and $s_{11}$ are scalars, $\omega_{21}$ and $s_{21}$ are $(N-1)\times 1$ vectors, and $\Omega_{22}$ and $S_{22}$ are $(N-1)\times(N-1)$ matrices. In each step of the block Gibbs sampler, we draw a diagonal element $\omega_{11}$ and off-diagonal elements $\omega_{21}$ from their full conditional posterior distributions. Without loss of generality, rows and columns of $\Omega$ can be rearranged so that the upper-left corner of $\Omega$, $\omega_{11}$, should be the diagonal element to be drawn from its full conditional posterior distribution. By using $\Omega$ and $S$ in \eqref{partition}, we have
\begin{align*}
    \mathrm{tr}\left(\Omega S\right) &= s_{11}\omega_{11} + 2s_{21}^{\intercal}\omega_{21} + \mathrm{tr}\left(\Omega_{22} S_{22}\right),
\end{align*}
and
\begin{align*}
    \left|\Omega\right| &= \left|\omega_{11}-\omega_{21}^{\intercal}\Omega_{22}^{-1}\omega_{21}\right|\left|\Omega_{22}\right|.
 \end{align*}
Then \eqref{sn.likelihood2} is rewritten as
\begin{align}
    \label{likelihood.trace.form1}
    p(R|\mu,\delta,\Omega,Z)
    &\propto |\Omega|^{\frac{T}2}\exp\left[-\frac12\mathrm{tr}(\Omega S)\right] \nonumber \\
    &\propto \left|\omega_{11}-\omega_{21}^{\intercal}\Omega_{22}^{-1}\omega_{21}\right|^{\frac{T}2} \left|\Omega_{22}\right|^{\frac{T}2} \nonumber \\
    &\quad \times\exp\left[-\frac12\left\{s_{11}\omega_{11} + 2s_{21}^{\intercal}\omega_{21} + \mathrm{tr}\left(\Omega_{22}S_{22}\right)\right\}\right].
\end{align}
Furthermore, following \cite{Wang}, we reparameterize $(\omega_{11},\omega_{21})$ to $(\eta,\omega_{21})$ where
\[
    \eta = \omega_{11}-\omega_{21}^{\intercal}\Omega_{22}^{-1}\omega_{21}.
\]
Finally we have
\begin{align}
    \label{likelihood.trace.form2}
    p(R|\mu,\delta,\Omega,Z)
    &\propto \eta^{\frac{T}2}\exp\left[-\frac12\left\{s_{11}\eta + s_{11}\omega_{21}^{\intercal}\Omega_{22}^{-1}\omega_{21} + 2s_{21}^{\intercal}\omega_{21}\right\}\right],
\end{align}
where we ignore the parts that do not depend on $\eta$ nor $\omega_{21}$.

We need to be careful in choosing the prior distribution of $(\eta,\omega_{21})$. Given that $\Omega$ from the previous iteration of the block Gibbs sampler is positive definite, newly generated $\omega_{11}$ and $\omega_{21}$ must satisfy
\begin{equation}
    \label{zyoken}
    \omega_{11} > \omega_{21}^{\intercal}\Omega_{22}^{-1}\omega_{21},
\end{equation}
to ensure that the updated $\Omega$ is also positive definite.
This condition \eqref{zyoken} requires
\[
    \eta =  \omega_{11} - \omega_{21}^{\intercal}\Omega_{22}^{-1}\omega_{21} > 0.
\]
Hence, we can use a gamma distribution:
\begin{equation}
    \label{eta.prior}
    \eta \sim Ga(a_\eta,b_\eta),
\end{equation}
as the prior distribution of $\eta$. Moreover, we suppose the prior distribution of off-diagonal elements $\omega_{21}$ is a truncated multivariate normal distribution:
\begin{equation}
    \label{omega21.prior}
    p(\omega_{21}|\omega_{11},\Omega_{22}) \propto \exp\left(-\frac12\omega_{21}^{\intercal}A_{\omega}\omega_{21}\right)\mathbf{1}_{M^+}(\omega_{21}),
\end{equation}
where
\[
    A_{\omega} = \frac1{\psi^2}\mathrm{diag}\left(\frac1{\rho_{12}^2},\dots,\frac1{\rho_{1N}^2}\right),\quad
    M^+ = \{\omega_{21}: \omega_{11} < \omega_{21}^{\intercal}\Omega_{22}^{-1}\omega_{21}\},
\]
in order to assure that the condition \eqref{zyoken} holds in the course of sampling. Applying Bayes' theorem to \eqref{eta.prior} and \eqref{likelihood.trace.form2}, we have
\begin{equation}
    \label{fc.eta}
    \eta|\cdot \sim Ga\left(a_\eta + \frac{T}2, b_\eta + \frac{s_{11}}2\right).
\end{equation}
With \eqref{omega21.prior}, \eqref{omega.horseshoe2} and \eqref{likelihood.trace.form2}, the full conditional posterior distribution of $\omega_{21}$ is derived as
\begin{equation}
    \label{fc.omega21}
    \omega_{21}|\cdot \sim \mathcal{N}\left(-\hat A_{\omega}^{-1}s_{21},\ \hat A_{\omega}^{-1}\right)\mathbf{1}_{M^+}(\omega_{21}),\quad \hat A_{\omega} = A_{\omega} + s_{11}\Omega_{22}^{-1}.
\end{equation}
In order to draw $\omega_{21}$ from \eqref{fc.omega21}, we apply the Hit-and-Run algorithm (\cite{Beslie}) as in \cite{ONGLASSO}.
\begin{description}
 \item[Step 1:] Pick a point $\alpha$ on the unit sphere randomly as $\alpha=\frac{u}{\Vert u\Vert}$, $u\sim\mathcal{N}(0,I)$.
 \item[Step 2:] Draw a random scalar $\kappa$ from $\mathcal{N}\left(\mu_{\kappa},\sigma_{\kappa}^2\right)\mathbf{1}_{R^{+}}(\kappa)$ where
  \begin{align*}
    \mu_{\kappa} &= -\frac{s_{21}^{\intercal}\alpha+\omega_{21}^{\intercal}\hat A_{\omega}\alpha}{\alpha^{\intercal}\hat A_{\omega}\alpha},\quad \sigma_{\kappa}^2 = \frac1{\alpha^{\intercal}\hat A_{\omega}\alpha}, \\
    R^{+} &= \left\{\kappa : \frac{-b_{\kappa}-\sqrt{b_{\kappa}^2-a_{\kappa}c_{\kappa}}}{a_{\kappa}} < \kappa <  \frac{-b_{\kappa}+\sqrt{b_{\kappa}^2-a_{\kappa}c_{\kappa}}}{a_{\kappa}}\right\}, \\
     a_{\kappa} &= \alpha^{\intercal}\Omega_{22}^{-1}\alpha,\quad
     b_{\kappa} = \omega_{21}^{\intercal}\Omega_{22}^{-1}\alpha,\quad
     c_{\kappa} = \omega_{21}^{\intercal}\Omega_{22}^{-1}\omega_{21} - \omega_{11}.
  \end{align*}
 \item[Step 3:] Update the old $\omega_{21}$ with $\omega_{21}+\kappa\alpha$.
\end{description}

Finally it is straightforward to derive the full conditional posterior distributions of hyper-parameters and auxiliary variables:
\begin{align}
    \rho_{ij}^2|\cdot &\sim IG\left(1,\frac1{\upsilon_{ij}}+\frac{\omega_{ij}^2}{2\psi^2}\right),\quad (1\leqq i < j\leqq N), \\
    \psi^2|\cdot &\sim IG\left(\frac{N(N-1)}4+\frac12,\ \frac1{\zeta}+\frac12\sum_{j=2}^N\sum_{i=1}^{j-1}\frac{\omega_{ij}^2}{\rho_{ij}^2}\right), \\
    \upsilon_{ij}|\cdot &\sim IG\left(1,1+\frac1{\rho_{ij}^2}\right),\quad (1\leqq i < j\leqq N), \\
    \zeta|\cdot &\sim IG\left(1,1+\frac1{\psi^2}\right).
\end{align}

Although we will examine the skew-normal distribution in the next section in order to simply compare the \cite{Harvey2010}'s sampling method in terms of identification of $\Delta$, we can easily extend the multivariate skew-normal model \eqref{sn.def1} to the multivariate skew-t distribution. See Appendix.

\section{Performance Comparisons with Simulation}
\label{Simulation}
In this section, we report results of Monte Carlo experiments to compare three models (Full-NOWI, LT-NOWI and LT-HSGHS), which are summarized in Table \ref{table:model}, in terms of accuracy in the parameter estimation.

\begin{table}[htbp]
\caption{Overview of comparative models}
\label{table:model}
\begin{center}
\begin{footnotesize}
\begin{tabular}{lccc}\hline
  & Constraint for $\Delta$ & Prior for $\Delta$ & Prior for $\Omega$ \\
  \hline
  Full-NOWI & Nothing  & Normal & Wishart \\
  LT-NOWI & Positive Lower-Triangular & Normal & Wishart \\
  LT-HSGHS & Positive Lower-Triangular & Horseshoe & Graphical Horseshoe \\
  \hline
\end{tabular}
\end{footnotesize}
\end{center}
\end{table}

We assume the following three designs of $\Delta$ in this simulation:
\begin{enumerate}
\item $\Delta$-Diag: $\Delta_{ii}=2.0\quad(i = 2n-1)$,  $\Delta_{ii}=-2.0\quad(i = 2n)$, otherwise 0.0.
\item $\Delta$-Sparse: $\Delta_{ii}=2.0\quad(i = 2n-1)$,  $\Delta_{ii}=-2.0\quad(i = 2n)$ , $\Delta_{i,i-1}=-1.0$, otherwise $0.0$.
\item $\Delta$-Dense:  $\Delta_{ii}=2.0\quad(i = 2n-1)$,  $\Delta_{ii}=-2.0\quad(i = 2n)$, $\Delta_{i,i-1}=-1.0$, otherwise the elements in lower-triangular equals to 1.0.  
\end{enumerate}
Since our main purpose is to compare estimation of $\Delta$, we set a simple assumption for the other parameter; $\Omega$ is the identity matrix, $\mu$ is fixed to zero. We generate artificial data of t = 1,500 and n = 15 from the multivariate skew-elliptical distribution with each specification and evaluate the posterior statistics of each parameter via MCMC. The hyper-parameters in the prior distributions are set up as follows.
\begin{description}
  \item[Full-NOWI] $b_\mu=0$, $A_\mu = 0.01I$,  $b_\Delta=0$, $A_\Delta = 0.01I$, $S_\Omega=nI$, $\nu_\Omega=n$ in \eqref{prior0}.
  \item[LT-NOWI] $b_\mu=0$, $A_\mu = 0.01I$,  $b_\delta=0$, $A_\delta = 0.01I$, $S_\Omega=nI$, $\nu_\Omega=n$ in \eqref{prior}.
  \item[LT-HSGHS] $b_\mu=0$, $A_\mu = 0.01I$,  $b_\delta=0$, $A_\delta = 0.01I$ in \eqref{prior}; $a_\eta=1.0$ and $b_\eta=0.0$ in \eqref{eta.prior}; $A_{\omega}=0.01I$ in \eqref{omega21.prior}
\end{description}
In all cases, the number of burn-in iterations were 50,000, and the Monte Carlo sample from the following 100,000 iterations was used in the Bayesian inference. Also, we repeated simulations 30 times for each setup and obtained a set of point estimates of $\Delta$ and $\Omega$. All computations are implemented with Python 3.7.0 on a desktop PC with 128GB RAM, 8GB GPU and eight-core 3.8GHz i7-10700K Intel processor.

To compare the three models in terms of accuracy in the point estimation of $\Delta$ and $\Omega$, we computed the Frobenius norm, as measurement of discrepancy between the point estimate and the true structure. Table 2 and 3 show the sample median loss with 30 replications for three models. The figures in parentheses are the standard errors. The smaller the value of the Frobenious norm, the closer the estimated structure is to the true one. In addition, in order to make the estimation results visually easy to understand, the posterior averages of $\Delta$ and $\Omega$ of each model in the 30th replication are shown in Figures 1 -- 6.

First, regarding $\Delta$, the Frobenious norm of the proposed models (LT-NOWI, LT-HSGHS) have decreased to 1/8 or less of the Full-NOWI model for all designs and the estimation accuracy has remarkably improved in Table 2. This is because the columns of $\Delta$ is not identified at all in Full-NOWI. On the other hand, this identification issue is resolved in LT-NOWI and LT-HSGHS and the structure of $\Delta$ can be estimated well with the proposed method as shown in Figures 1, 3 and 5. Furthermore, for the $\Delta$-Diag case and the $\Delta$-Sparse case, Table 2 reports that the Frobenious norm of LT-HSGHS is less than half the value of LT-NOWI. This is because that the horseshoe prior in LT-HSGHS contributes to the estimation performance by shrinking non-essential elements to zero, while a large amount of non-zero entries still remain in $\Delta$ for LT-NOWI as shown in Figures 1 and 3. However, the difference in the Frobenious norm between LT-NOWI and LT-HSGHS becomes smaller in the $\Delta$-Dense design because the sparse assumption of $\Delta$ is not satisfied in this case.

Next, let take a look at results on $\Omega$. Note that the true structure of $\Omega$ is the identity matrix. We examine how the estimation accuracy of $\Omega$ changes across the structural designs of $\Delta$. For all designs of $\Delta$, the estimation accuracy is significantly improved in LT-HSGHS, the value of Frobenious norm is 1/3 or less in Table 3 compared with Full-NOWI and LT-NOWI. In fact, there are a lot of non-zero entries in the off-diagonal elements in Full-NOWI and LT-NOWI in Figures 2, 4 and 6. On the other hand, the posterior mean of $\Omega$ in LT-HSGHS becomes the diagonal matrix thanks to the shrinkage effect. Also, comparing LT-NOWI with Full-NOWI, the Frobenious norm is slightly smaller in LT-NOWI for all $\Delta$ designs. These findings suggest that the posterior distribution of $\Omega$ is affected by the estimation of $\Delta$ as shown in \eqref{fc.omega}.

\begin{table}[htbp]
\caption{Sample median loss in the point estimation of $\boldsymbol{\Delta}$}
\begin{center}
\begin{footnotesize}
\begin{tabular}{lccc}\hline
    & $\Delta$-Diag & $\Delta$-Sparse & $\Delta$-Dense\\ \hline
    & & &  \\
\multicolumn{2}{l}{\underline{Frobenius norm}} \\
    & & &  \\
Full-NOWI & 10.587  & 11.588 & 12.620  \\
    &  (0.534) & (0.743) & (0.723)  \\ 
LT-NOWI & 1.344 & 1.362 & 1.420  \\
    &  (0.083)   & (0.115) & (0.108)  \\   
LT-HSGHS & \textbf{0.380} & \textbf{0.617} & \textbf{1.214}  \\
    &  (0.070)   & (0.075) & (0.138)  \\ \hline
Notes: & \multicolumn{2}{l}{(a) The smaller losses are boldfaced.} \\
        & \multicolumn{2}{l}{(b) The figures in parentheses are the standard errors.} \\
\end{tabular}
\end{footnotesize}
\end{center}
\end{table}

\begin{table}[htbp]
\caption{Sample median loss in the point estimation of $\boldsymbol{\Omega}$}
\begin{center}
\begin{footnotesize}
\begin{tabular}{lccc}\hline
    & $\Delta$-Diag & $\Delta$-Sparse & $\Delta$-Dense\\ \hline
    & & &  \\
\multicolumn{2}{l}{\underline{Frobenius norm}} \\
    & & &  \\
Full-NOWI & 2.550  & 2.447 & 2.352  \\
    &  (0.139) & (0.138) & (0.136)  \\ 
LT-NOWI & 2.255 & 2.238 & 2.210  \\
    &  (0.179)   & (0.147) & (0.155)  \\   
LT-HSGHS & \textbf{0.393} & \textbf{0.479} & \textbf{0.724}  \\
    &  (0.101)   & (0.160) & (0.288)  \\ \hline
Notes: & \multicolumn{2}{l}{(a) The smaller losses are boldfaced.} \\
        & \multicolumn{2}{l}{(b) The figures in parentheses are the standard errors.} \\
\end{tabular}
\end{footnotesize}
\end{center}
\end{table}

\begin{figure}[htbp]
\begin{tabular}{ccc}
\begin{minipage}[b]{0.5\hsize}
\centering
\includegraphics[width=6.5cm]{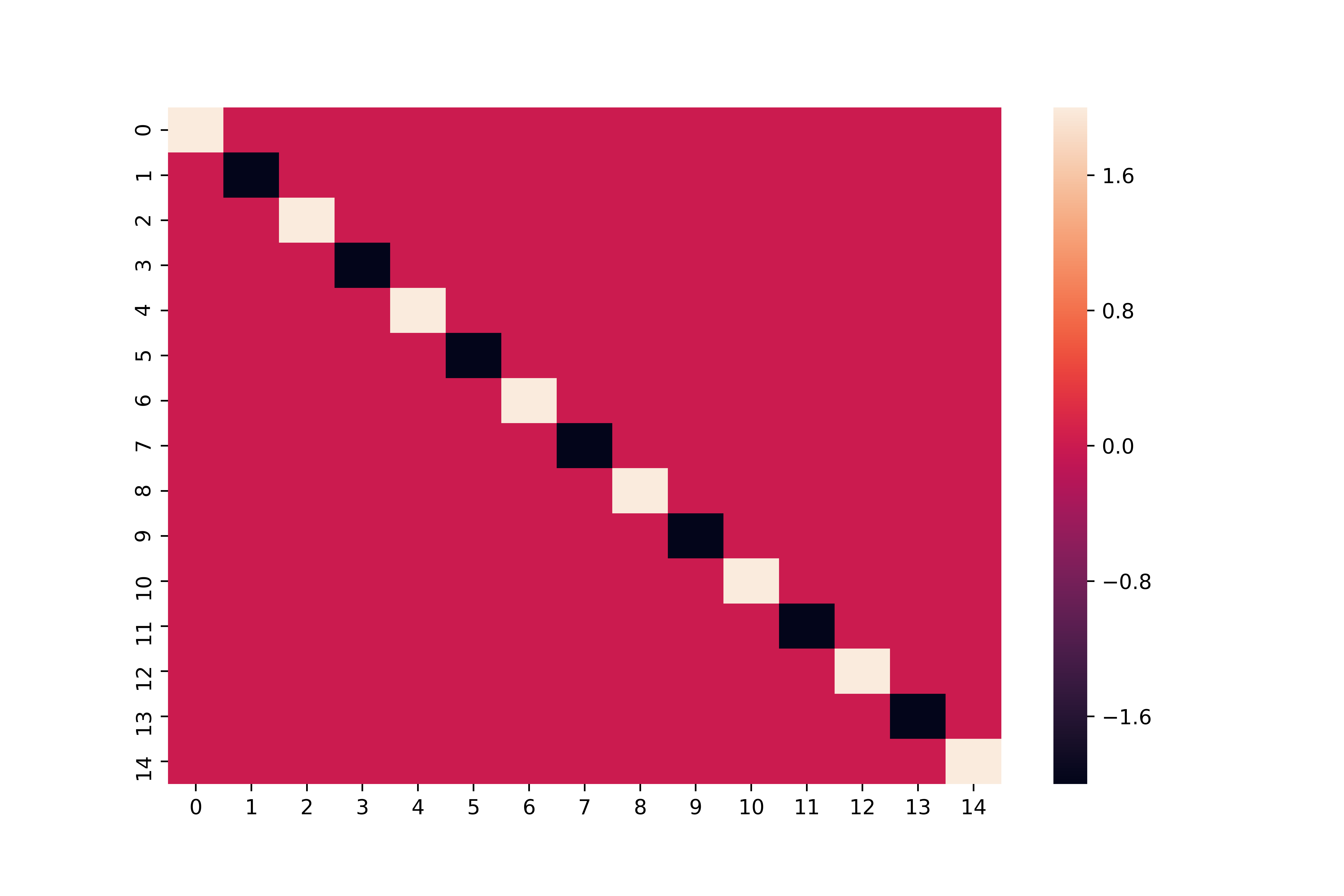}
\subcaption{True}
\end{minipage} &
\begin{minipage}[b]{0.5\hsize}
\centering
\includegraphics[width=6.5cm]{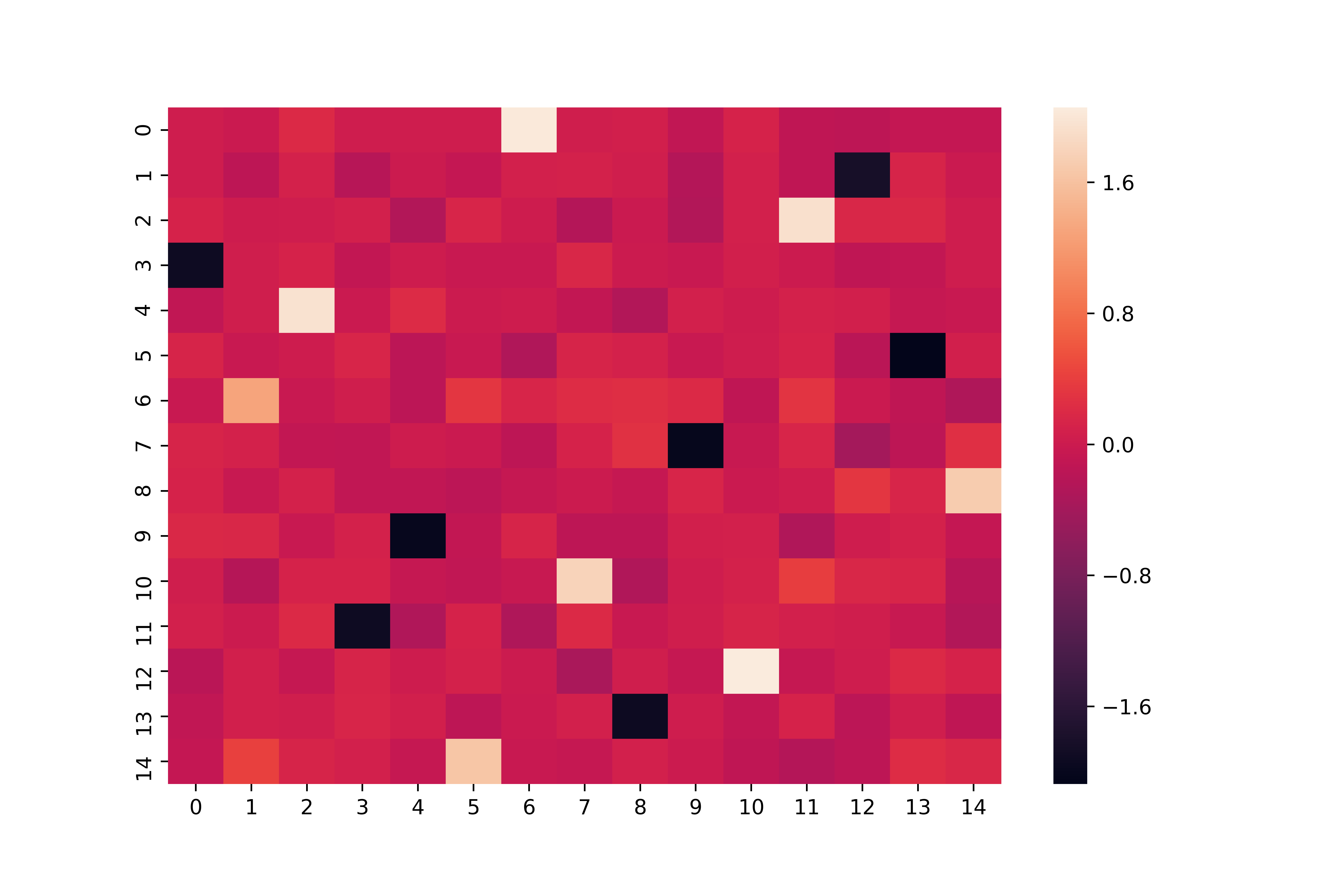}
\subcaption{Full-NOWI}
\end{minipage} \\
\begin{minipage}[b]{0.5\hsize}
\centering
\includegraphics[width=6.5cm]{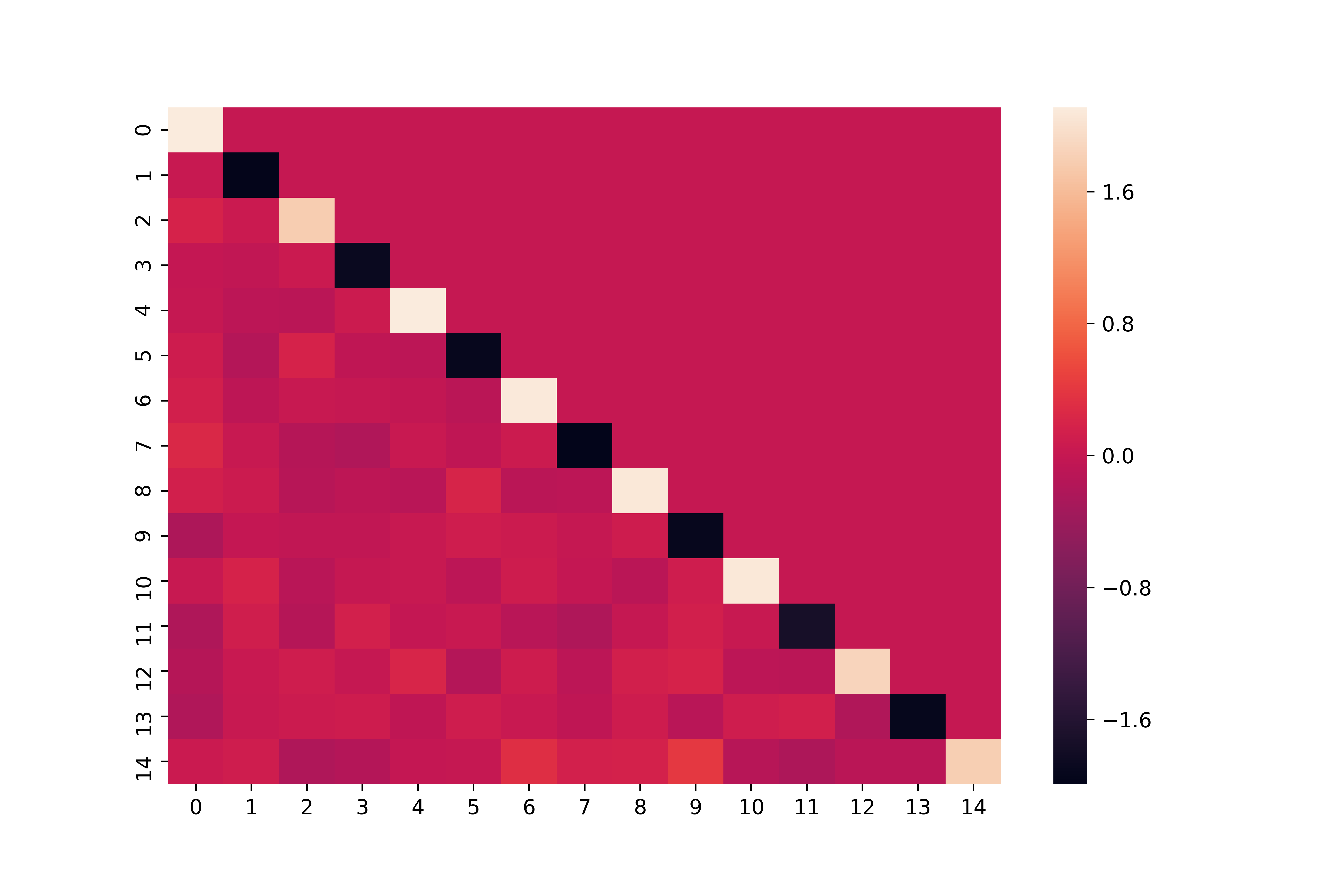}
\subcaption{LT-NOWI}
\end{minipage} &
\begin{minipage}[b]{0.5\hsize}
\centering
\includegraphics[width=6.5cm]{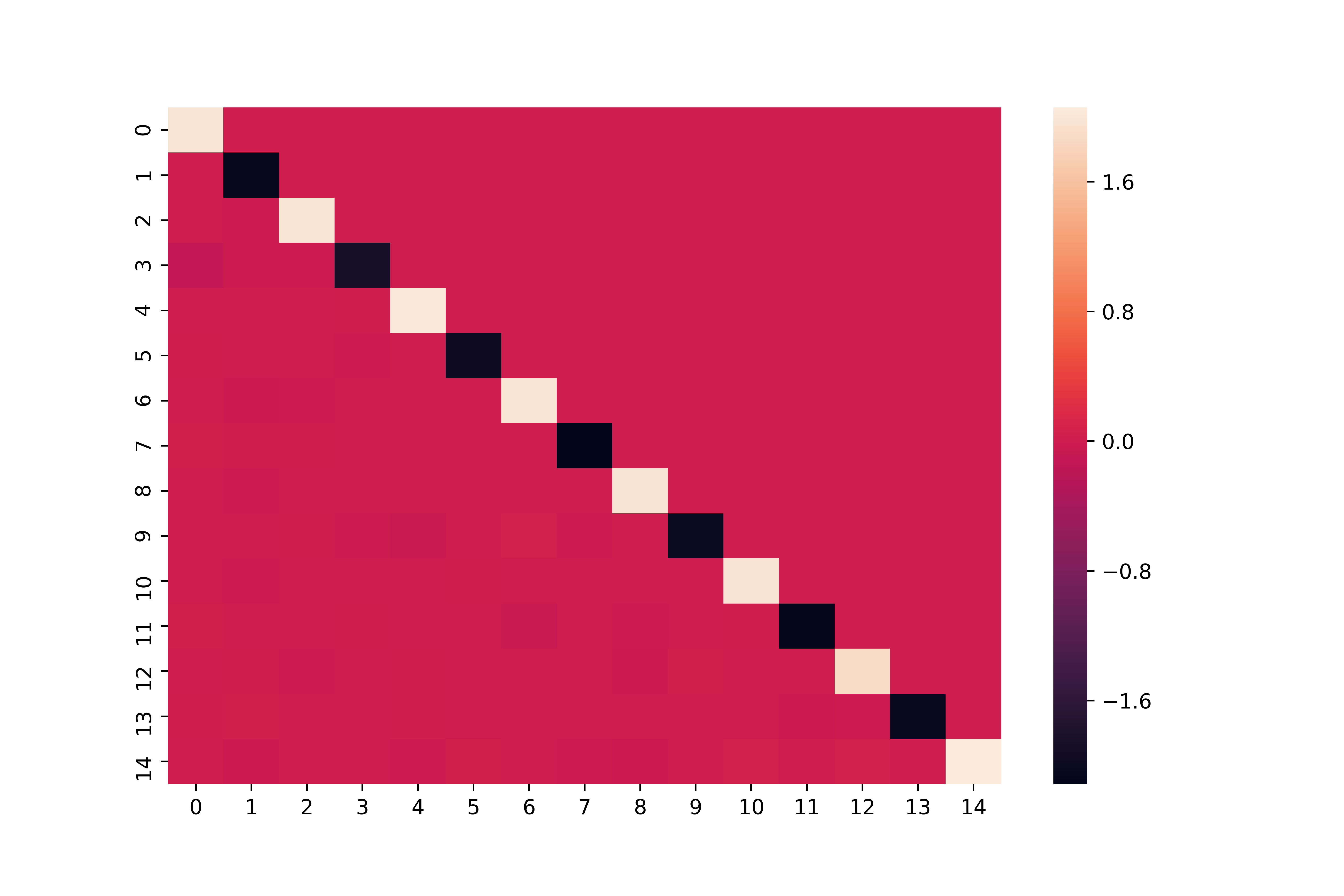}
\subcaption{LT-HSGHS}
\end{minipage}
\end{tabular}
\caption{$\Delta$-Diag: True Structure of $\Delta$ and estimated $\Delta$}
\end{figure}

\begin{figure}[htbp]
\begin{tabular}{ccc}
\begin{minipage}[b]{0.5\hsize}
\centering
\includegraphics[width=6.5cm]{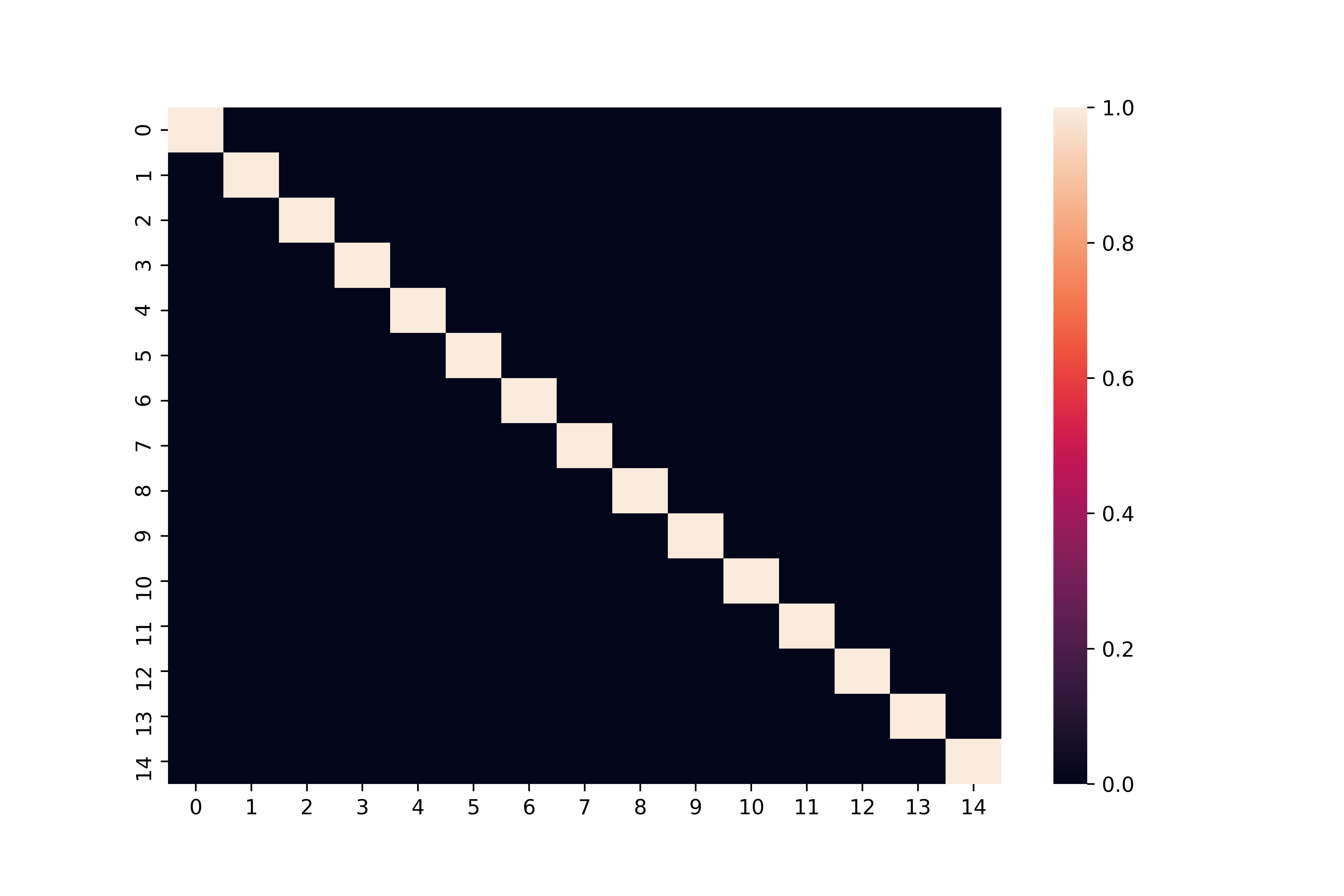}
\subcaption{True}
\end{minipage} &
\begin{minipage}[b]{0.5\hsize}
\centering
\includegraphics[width=6.5cm]{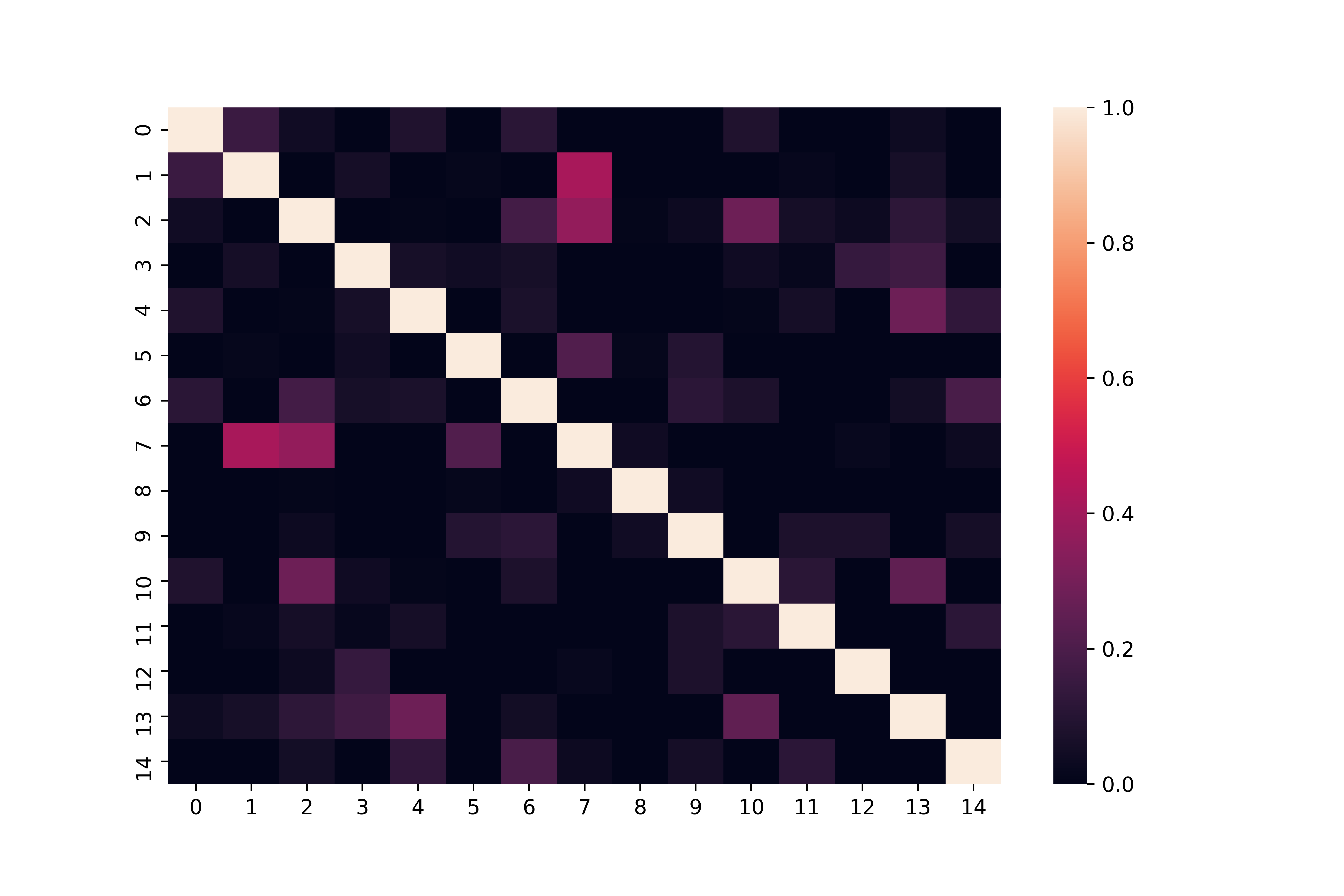}
\subcaption{Full-NOWI}
\end{minipage} \\
\begin{minipage}[b]{0.5\hsize}
\centering
\includegraphics[width=6.5cm]{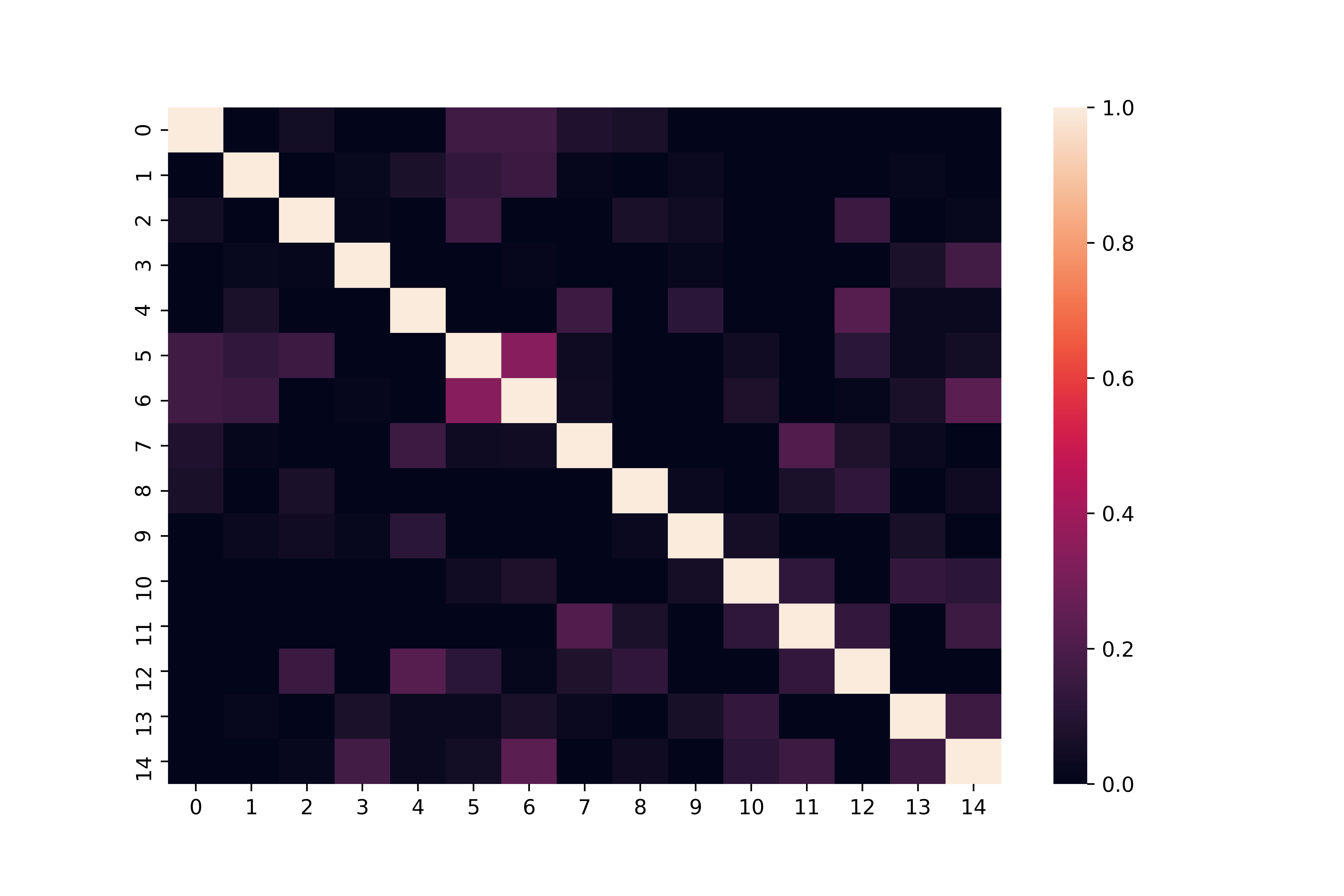}
\subcaption{LT-NOWI}
\end{minipage} &
\begin{minipage}[b]{0.5\hsize}
\centering
\includegraphics[width=6.5cm]{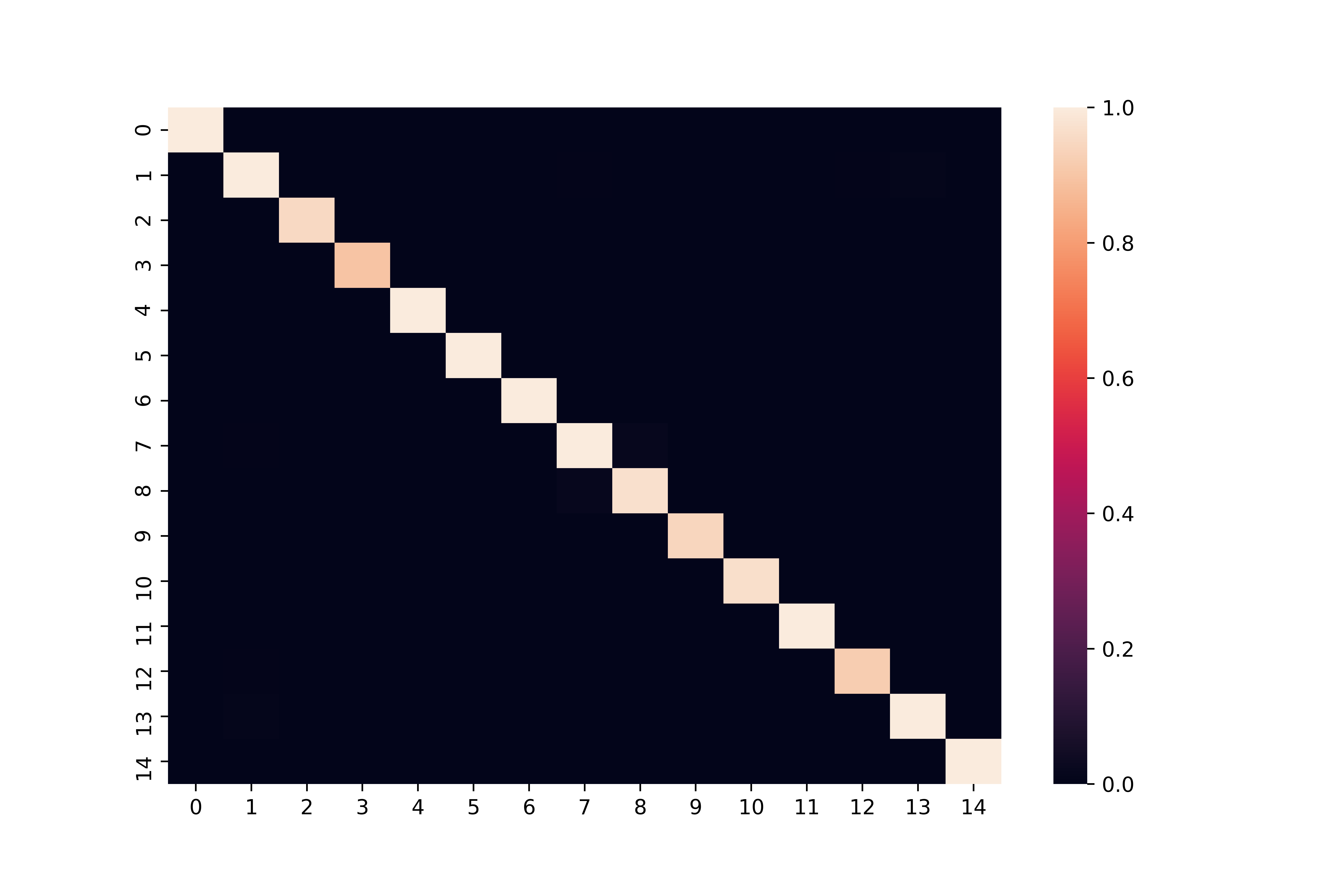}
\subcaption{LT-HSGHS}
\end{minipage}
\end{tabular}
\caption{$\Delta$-Diag: True Structure of $\Omega$ and estimated $\Omega$}
\end{figure}

\begin{figure}[htbp]
\begin{tabular}{ccc}
\begin{minipage}[b]{0.5\hsize}
\centering
\includegraphics[width=6.5cm]{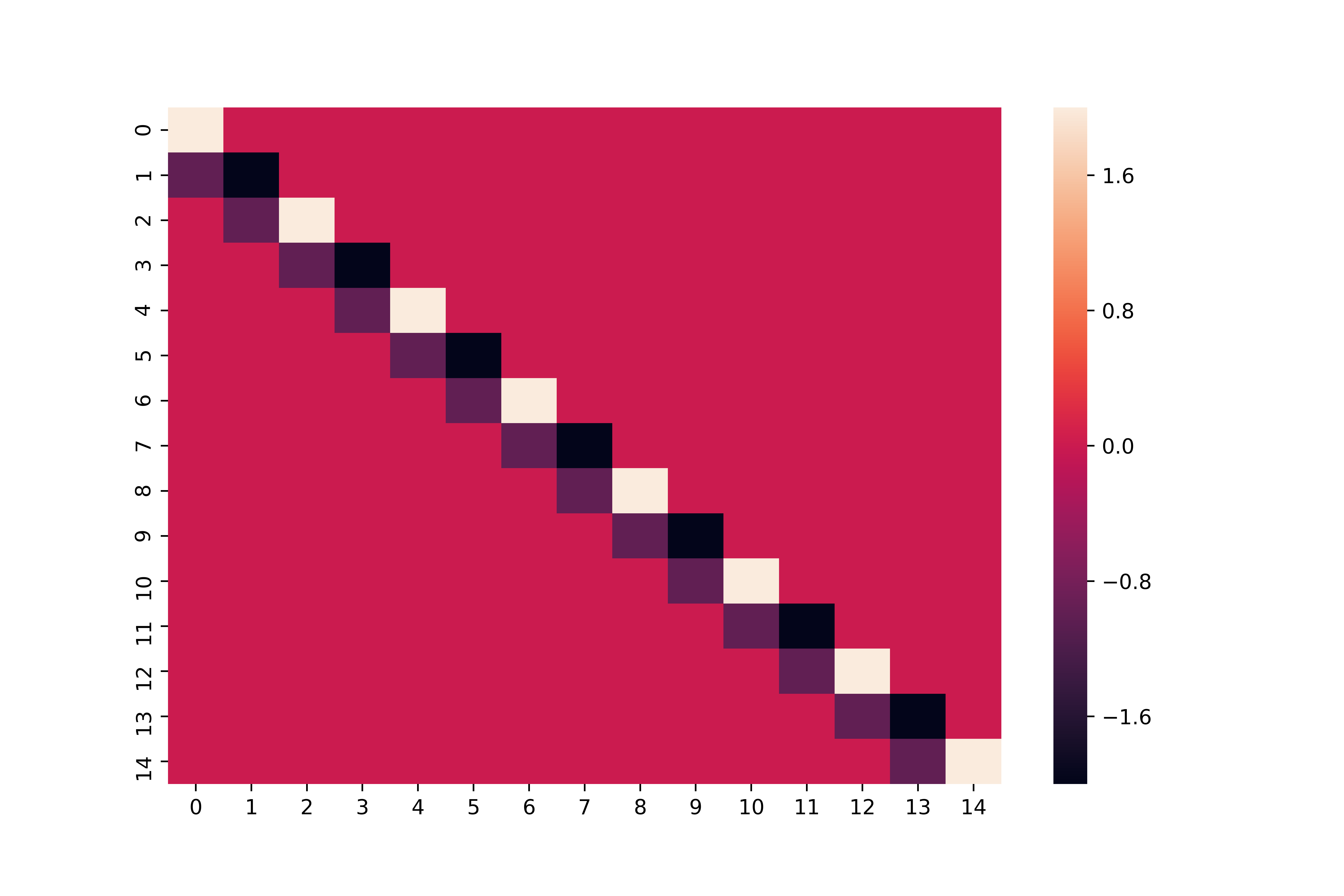}
\subcaption{True}
\end{minipage} &
\begin{minipage}[b]{0.5\hsize}
\centering
\includegraphics[width=6.5cm]{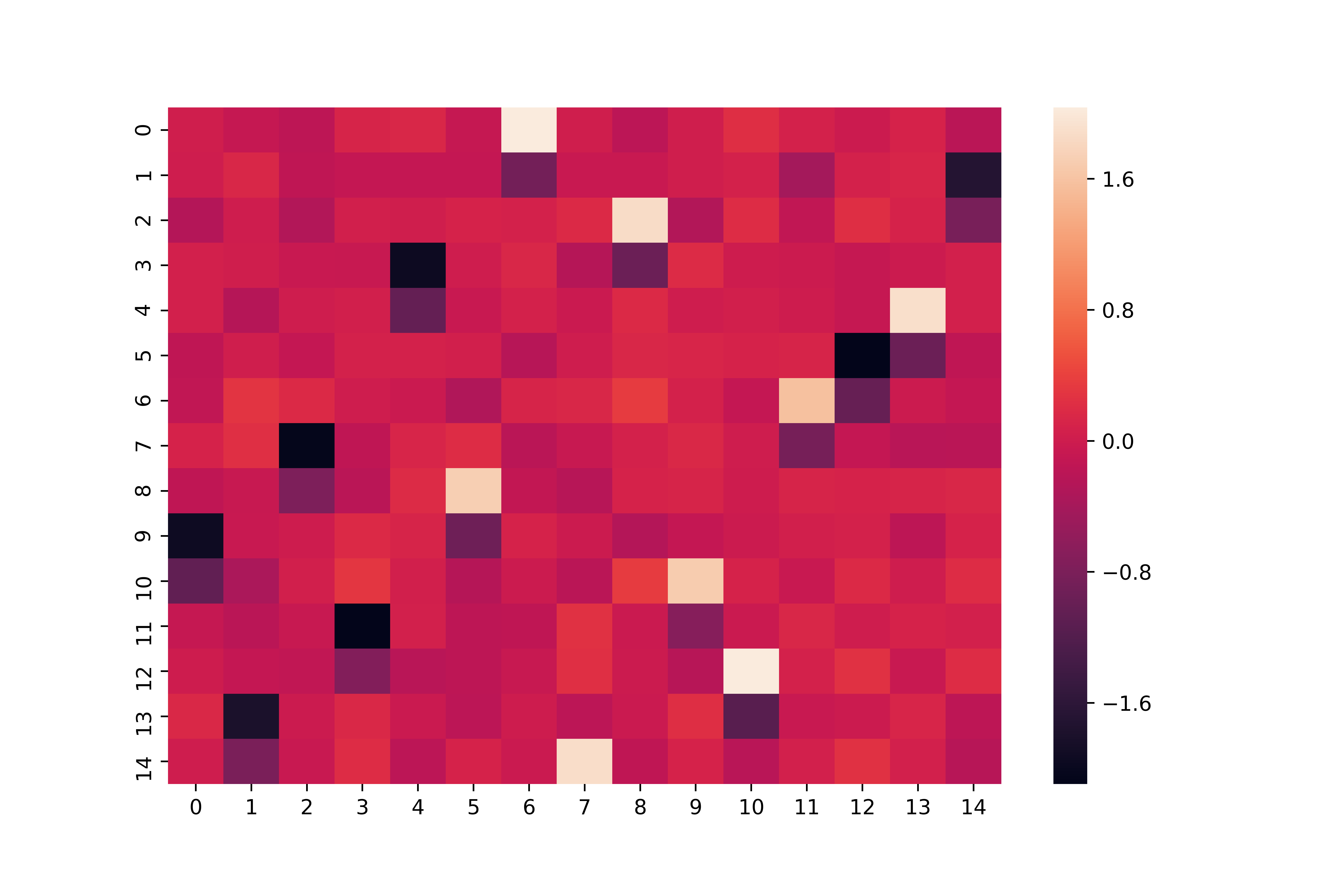}
\subcaption{Full-NOWI}
\end{minipage} \\
\begin{minipage}[b]{0.5\hsize}
\centering
\includegraphics[width=6.5cm]{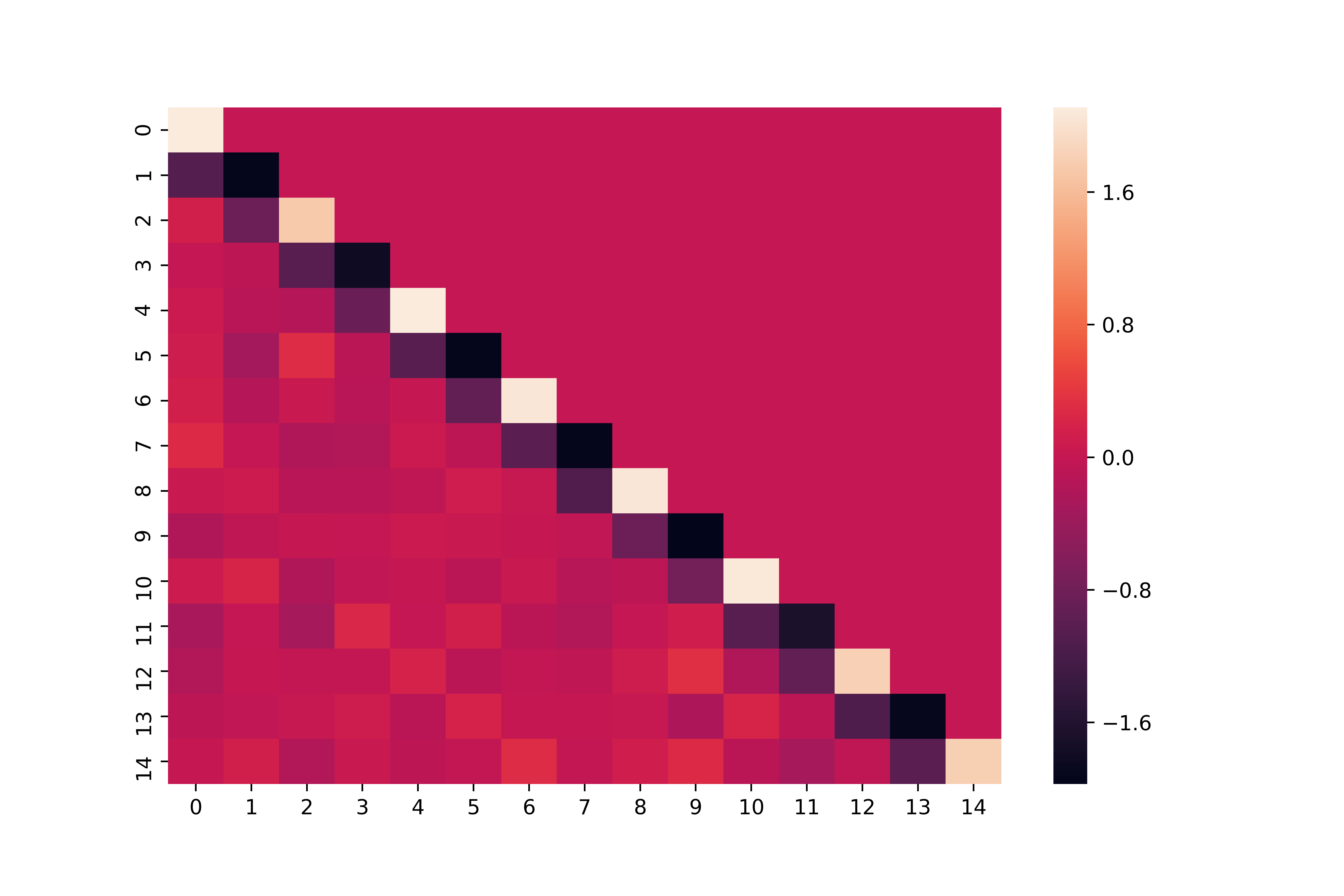}
\subcaption{LT-NOWI}
\end{minipage} &
\begin{minipage}[b]{0.5\hsize}
\centering
\includegraphics[width=6.5cm]{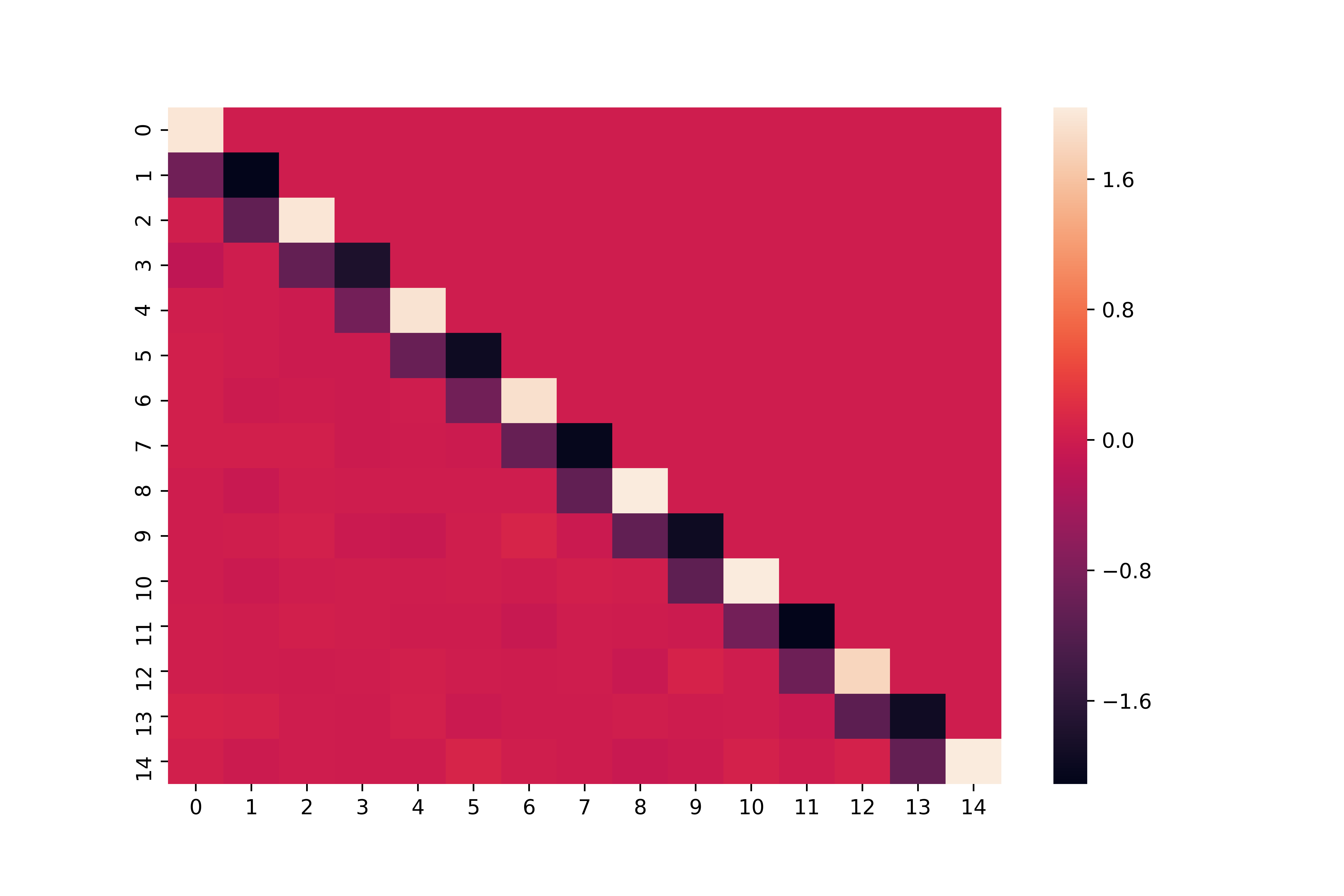}
\subcaption{LT-HSGHS}
\end{minipage}
\end{tabular}
\caption{$\Delta$-Sparse: True Structure of $\Delta$ and estimated $\Delta$}
\end{figure}

\begin{figure}[htbp]
\begin{tabular}{ccc}
\begin{minipage}[b]{0.5\hsize}
\centering
\includegraphics[width=6.5cm]{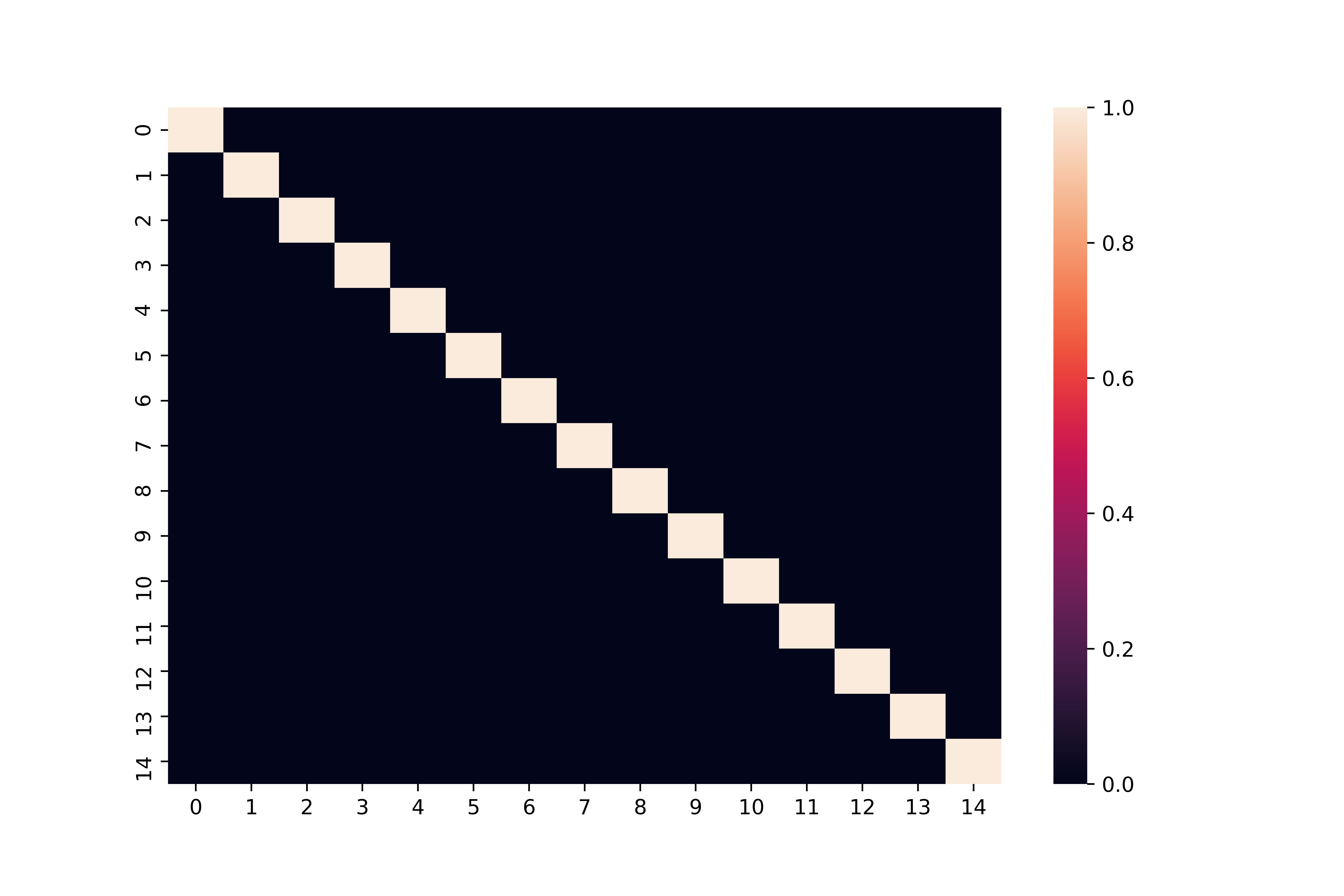}
\subcaption{True}
\end{minipage} &
\begin{minipage}[b]{0.5\hsize}
\centering
\includegraphics[width=6.5cm]{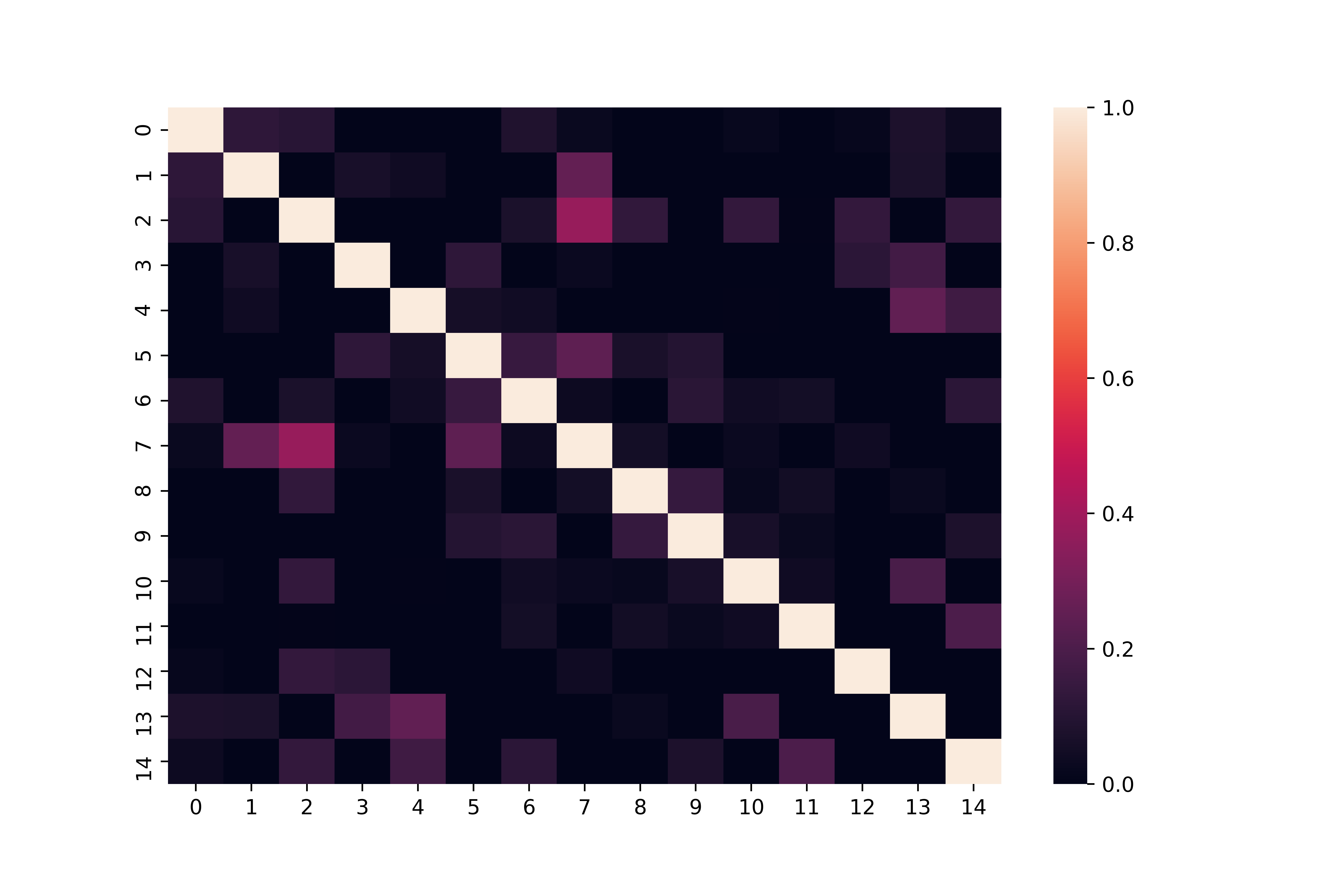}
\subcaption{Full-NOWI}
\end{minipage} \\
\begin{minipage}[b]{0.5\hsize}
\centering
\includegraphics[width=6.5cm]{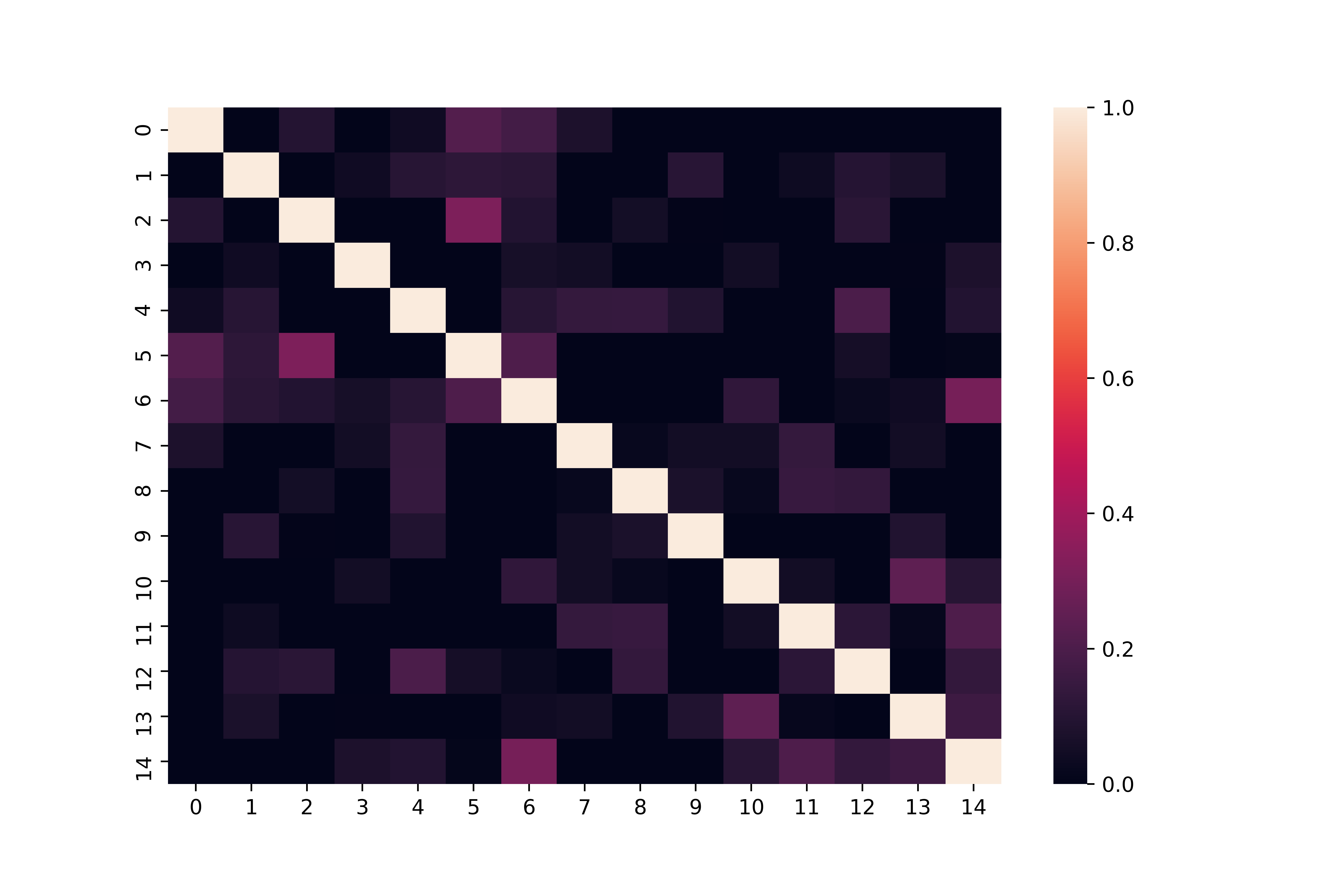}
\subcaption{LT-NOWI}
\end{minipage} &
\begin{minipage}[b]{0.5\hsize}
\centering
\includegraphics[width=6.5cm]{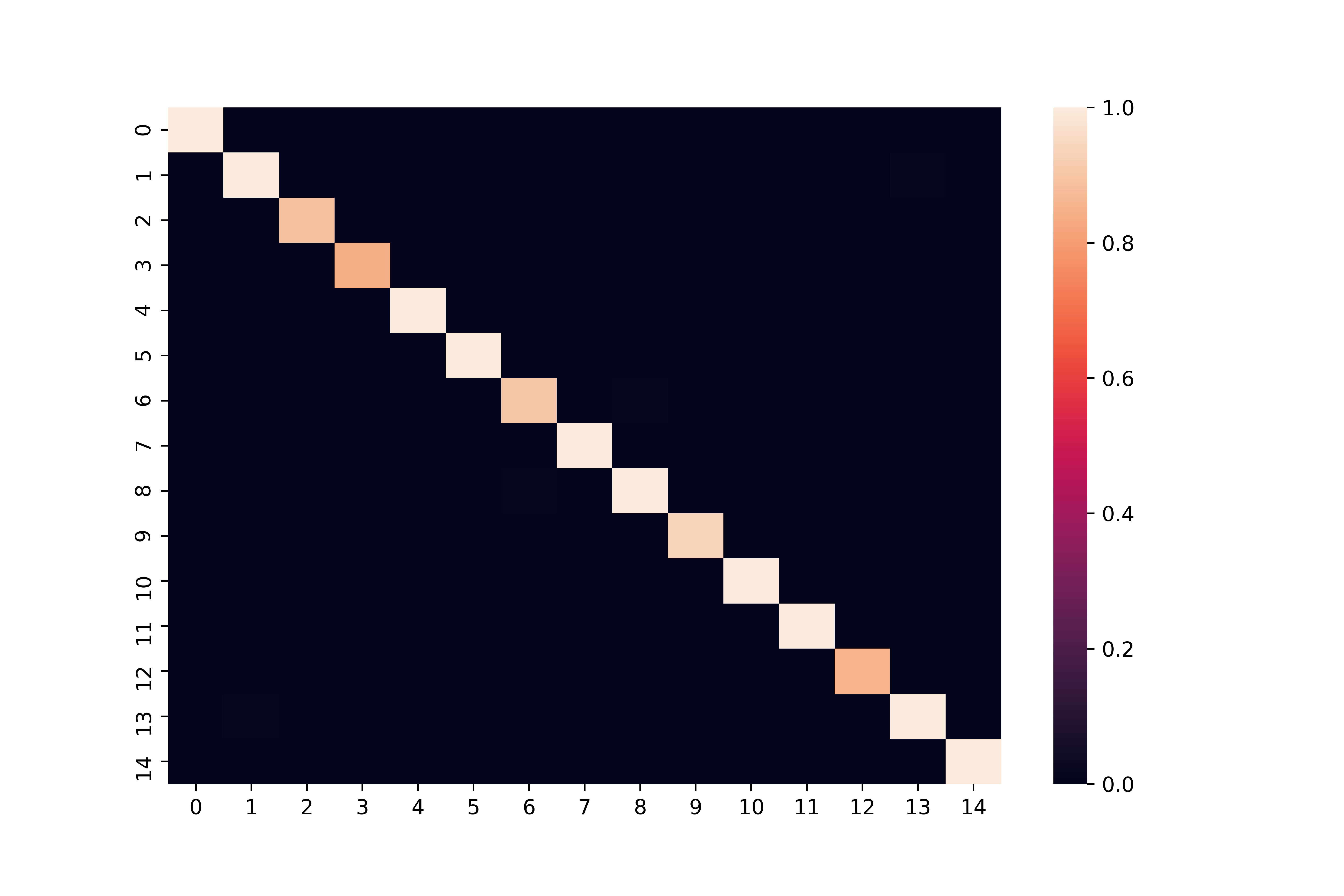}
\subcaption{LT-HSGHS}
\end{minipage}
\end{tabular}
\caption{$\Delta$-Sparse: True Structure of $\Omega$ and estimated $\Omega$}
\end{figure}

\begin{figure}[htbp]
\begin{tabular}{ccc}
\begin{minipage}[b]{0.5\hsize}
\centering
\includegraphics[width=6.5cm]{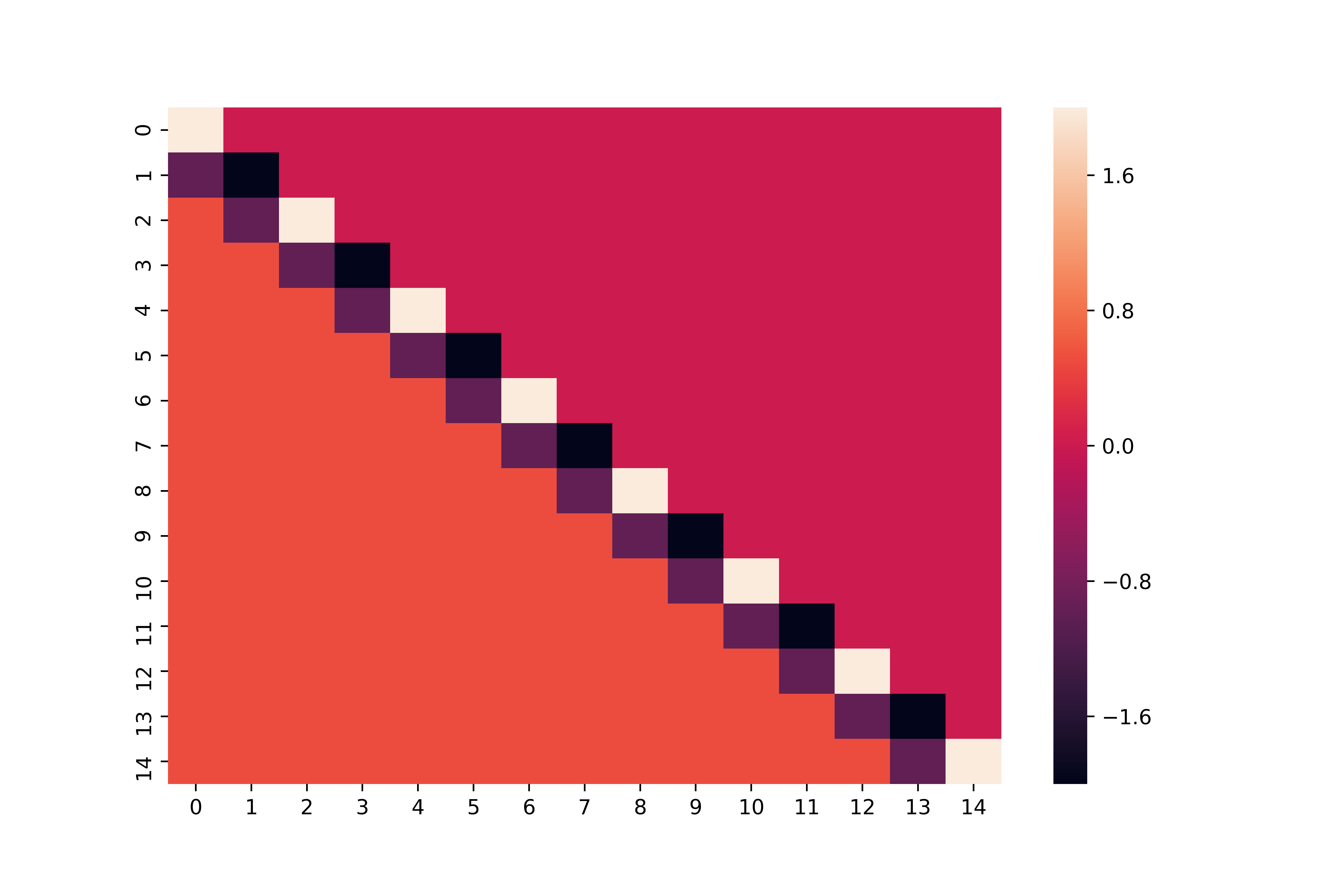}
\subcaption{True}
\end{minipage} &
\begin{minipage}[b]{0.5\hsize}
\centering
\includegraphics[width=6.5cm]{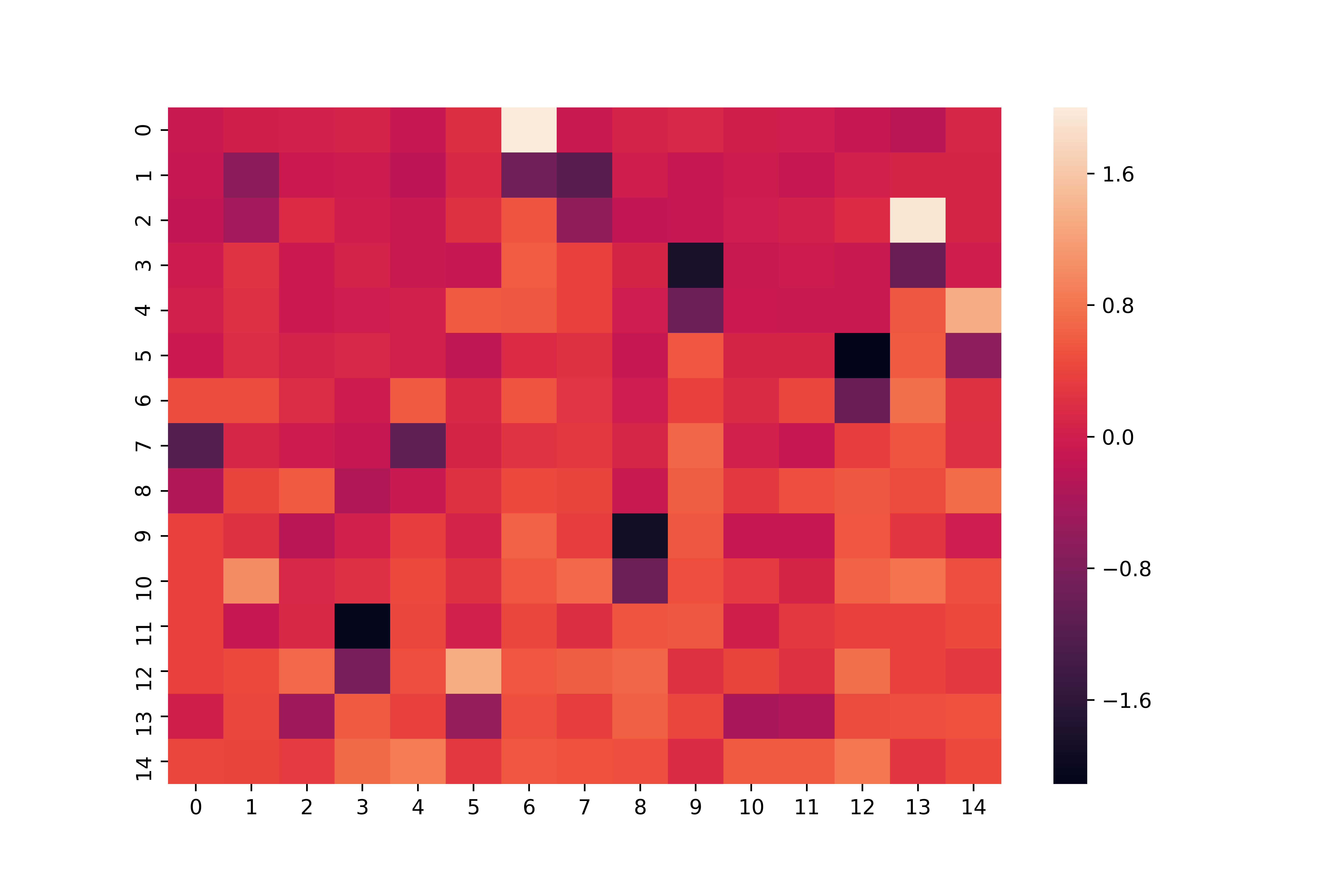}
\subcaption{Full-NOWI}
\end{minipage} \\
\begin{minipage}[b]{0.5\hsize}
\centering
\includegraphics[width=6.5cm]{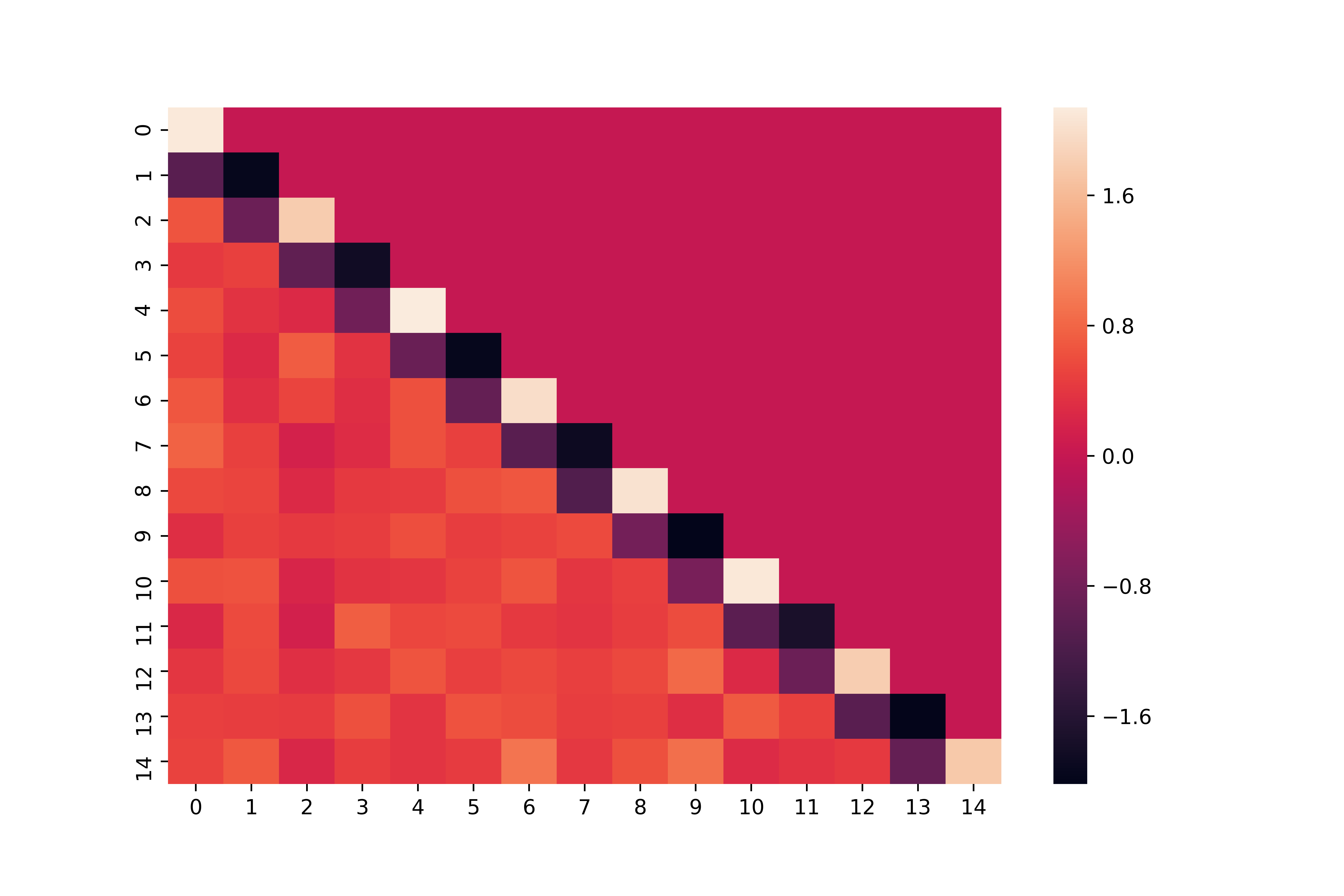}
\subcaption{LT-NOWI}
\end{minipage} &
\begin{minipage}[b]{0.5\hsize}
\centering
\includegraphics[width=6.5cm]{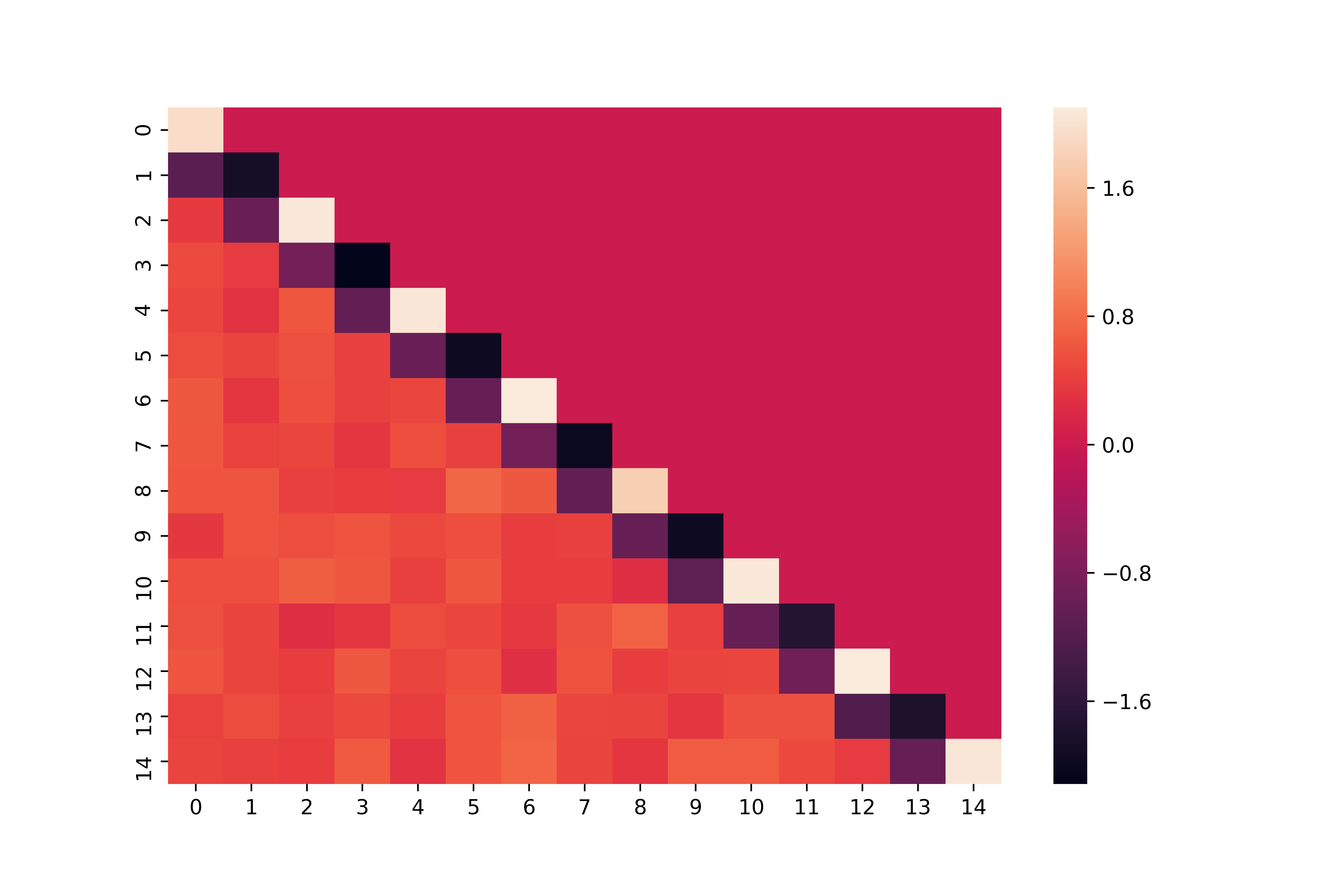}
\subcaption{LT-HSGHS}
\end{minipage}
\end{tabular}
\caption{$\Delta$-Dense: True Structure of $\Delta$ and estimated $\Delta$}
\end{figure}

\begin{figure}[htbp]
\begin{tabular}{ccc}
\begin{minipage}[b]{0.5\hsize}
\centering
\includegraphics[width=6.5cm]{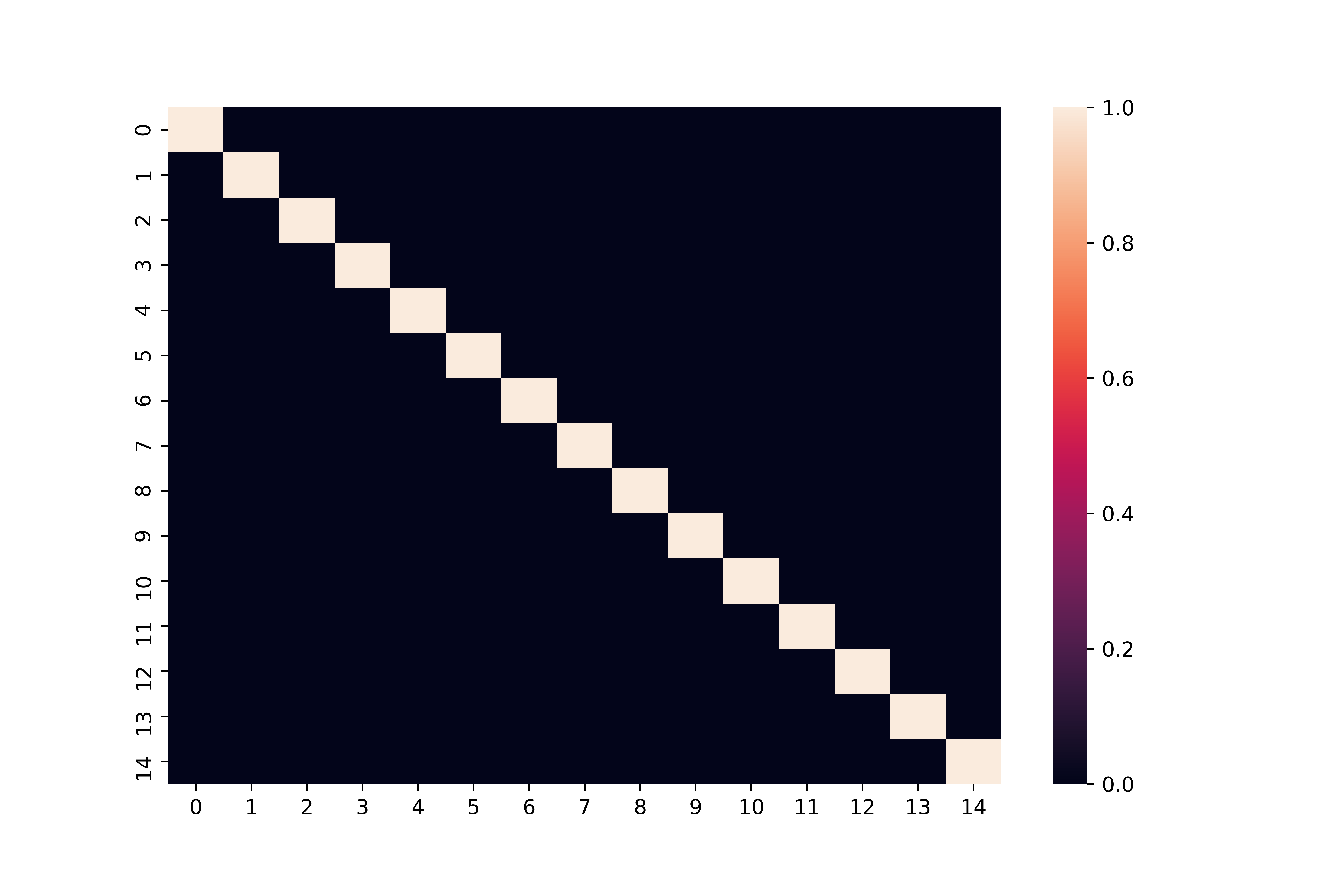}
\subcaption{True}
\end{minipage} &
\begin{minipage}[b]{0.5\hsize}
\centering
\includegraphics[width=6.5cm]{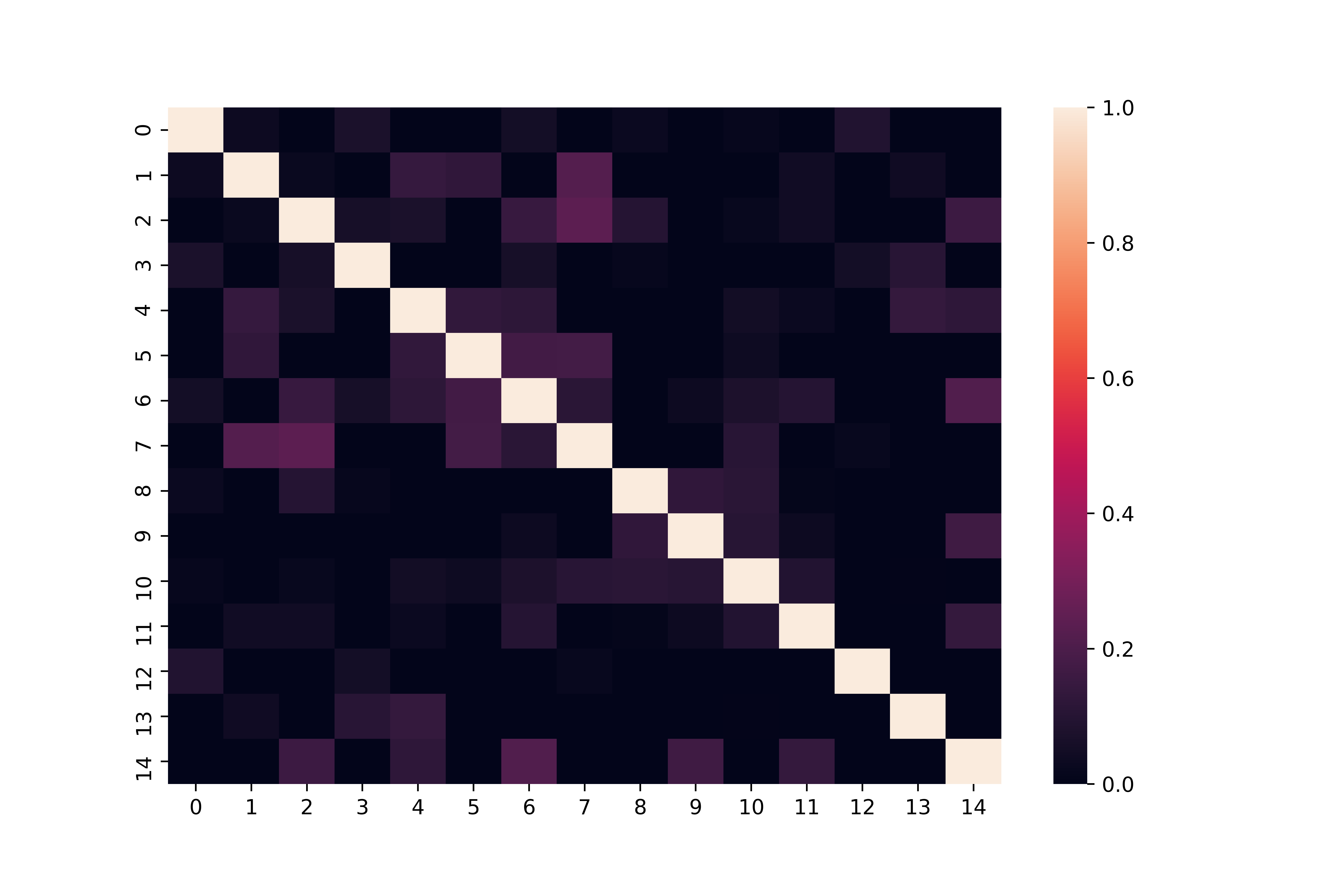}
\subcaption{Full-NOWI}
\end{minipage} \\
\begin{minipage}[b]{0.5\hsize}
\centering
\includegraphics[width=6.5cm]{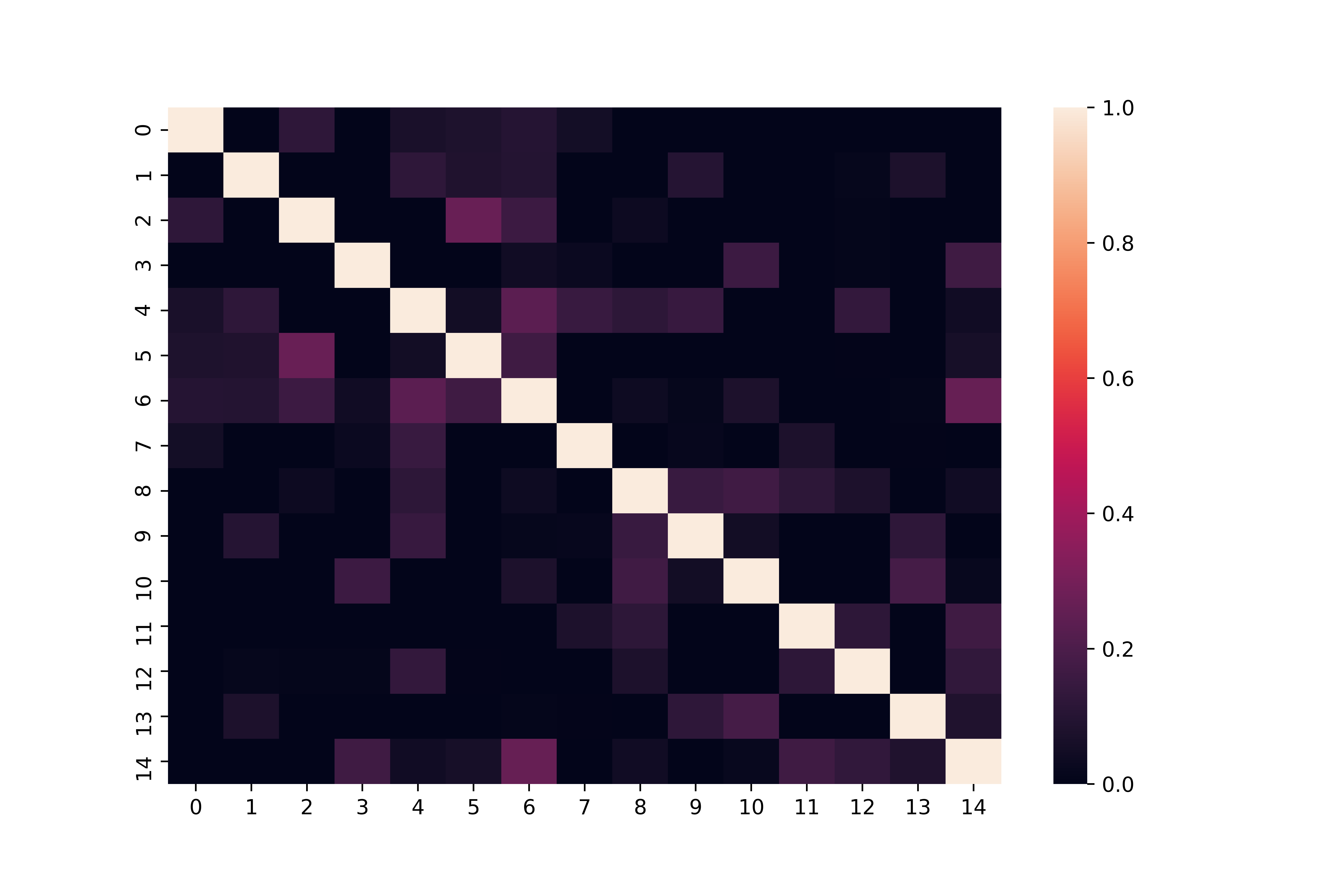}
\subcaption{LT-NOWI}
\end{minipage} &
\begin{minipage}[b]{0.5\hsize}
\centering
\includegraphics[width=6.5cm]{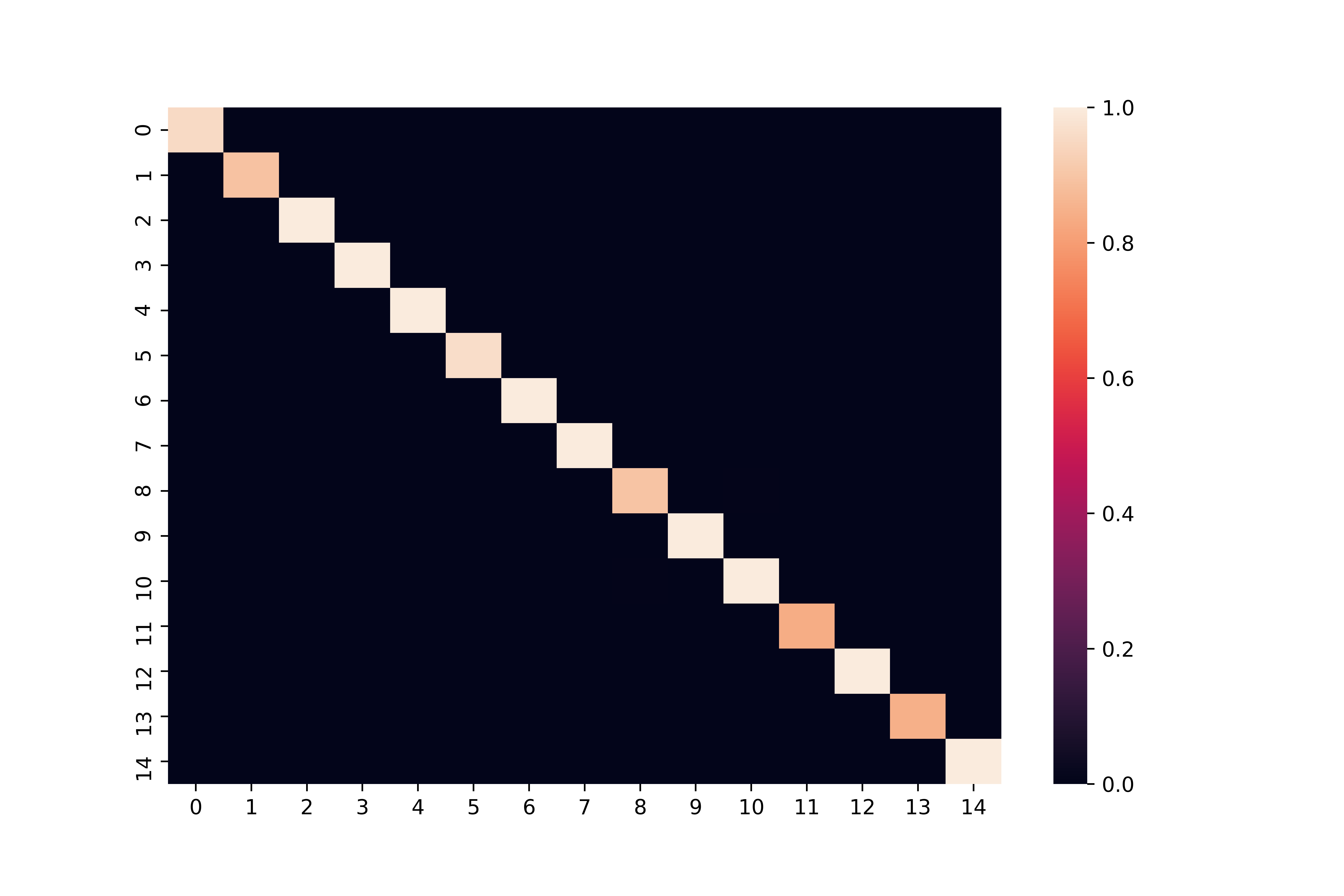}
\subcaption{LT-HSGHS}
\end{minipage}
\end{tabular}
\caption{$\Delta$-Dense: True Structure of $\Omega$ and estimated $\Omega$}
\end{figure}

\newpage
\section{Conclusion}
\label{conclusion}
In this paper, we have raised a possible identification issue on the skewness matrix of the skew-elliptical distribution in the Bayesian MCMC method proposed by \cite{Harvey2010} due to label switching. To avoid this issue, we proposed an modified model in which the lower-triangular constraint was imposed upon the skewness matrix. Moreover, we devised an extended model with the horseshoe prior for both skewness matrix and precision matrix to further improve the estimation accuracy.

In the simulation study, we compared the proposed models with the model of \cite{Harvey2010} in the three structural designs of the skewness matrix and found that the proposed models with the identification constraint significantly improved the estimation accuracy of the skewness matrix.

\section*{Acknowledgements}
This research is supported by the Keio University Doctorate Student Grant-in-Aid Program from Ushioda Memorial Fund; and JSPS KAKENHI under Grant [number MKK337J].

\section*{Conflict of interest}
The authors declare that they are funded by the Keio University Doctorate Student Grant-in-Aid Program from Ushioda Memorial Fund and JSPS KAKENHI under Grant [number MKK337J] to conduct this research.

\section*{Ethical standards}
The authors declare that the experiments in this paper comply with the current laws of Japan where we had conducted the experiment.

\bibliography{Skew_Elliptical_reference}

\begin{thebibliography}{32}
\providecommand{\natexlab}[1]{#1}
\providecommand{\url}[1]{\texttt{#1}}
\expandafter\ifx\csname urlstyle\endcsname\relax
  \providecommand{\doi}[1]{doi: #1}\else
  \providecommand{\doi}{doi: \begingroup \urlstyle{rm}\Url}\fi

\bibitem[Aas and Haff(2006)]{AasHaff}
K.~Aas and I.~H. Haff.
\newblock The generalized hyperbolic skew student's t-distribution.
\newblock \emph{Journal of Financial Econometrics}, 4\penalty0 (2):\penalty0
  275--309, 2006.

\bibitem[Adcock and Azzalini(2020)]{AdcockAzzalini2020}
C.~Adcock and A.~Azzalini.
\newblock A selective overview of skew-elliptical and related distributions and
  of their applications.
\newblock \emph{Symmetry}, 12\penalty0 (1), 2020.

\bibitem[Alodat and Al-Rawwash(2014)]{AR2014}
M.~T. Alodat and M.~Y. Al-Rawwash.
\newblock The extended skew gaussian process for regression.
\newblock \emph{METRON}, 72\penalty0 (3):\penalty0 317--330, 2014.

\bibitem[Azzalini and Capitanio(2003)]{Azzalini2003}
A.~Azzalini and A.~Capitanio.
\newblock Distributions generated by perturbation of symmetry with emphasis on
  a multivariate skew t-distribution.
\newblock \emph{Journal of the Royal Statistical Society: Series B (Statistical
  Methodology)}, 65\penalty0 (2):\penalty0 367--389, 2003.

\bibitem[Azzalini and Valle(1996)]{Azzalini96}
A.~Azzalini and D.~Valle.
\newblock The multivariate skew-normal distribution.
\newblock \emph{Biometrika}, 83\penalty0 (4):\penalty0 715--726, 1996.

\bibitem[Barbi and Romagnoli(2018)]{BarbiRomagnoli}
M.~Barbi and S.~Romagnoli.
\newblock Skewness, basis risk, and optimal futures demand.
\newblock \emph{International Review of Economics \& Finance}, 58:\penalty0
  14--29, 2018.

\bibitem[Barndorff-Nielsen(1977)]{BN77}
O.~E. Barndorff-Nielsen.
\newblock Exponentially decreasing distributions for the logarithm of particle
  size.
\newblock \emph{Proceedings of the Royal Society of London. A. Mathmatical and
  Physical Sciences}, 353:\penalty0 401--419, 1977.

\bibitem[B\'{e}lisle et~al.(1993)B\'{e}lisle, Romeijn, and Smith]{Beslie}
C.~J.~P. B\'{e}lisle, H.~E. Romeijn, and R.~L. Smith.
\newblock Hit-and-run algorithms for generating multivariate distributions.
\newblock \emph{Mathmatics of Operations Research}, 18\penalty0 (2):\penalty0
  255--266, 1993.

\bibitem[Branco and Dey(2001)]{BrancoDey}
M.~D. Branco and D.~K. Dey.
\newblock A general class of multivariate skew-elliptical distributions.
\newblock \emph{Journal of Multivariate Analysis}, 79\penalty0 (1):\penalty0
  99--113, 2001.

\bibitem[Carmichael and Co\"{e}n(2013)]{CarmichaelCoen}
B.~Carmichael and A.~Co\"{e}n.
\newblock Asset pricing with skewed-normal return.
\newblock \emph{Finance Research Letters}, 10\penalty0 (2):\penalty0 50--57,
  2013.

\bibitem[Carvalho et~al.(2010)Carvalho, Polson, and Scott]{Carvalho2010}
C.~M. Carvalho, N.~G. Polson, and J.~G. Scott.
\newblock The horseshoe estimator for sparse signals.
\newblock \emph{Biometrika}, 97\penalty0 (2):\penalty0 465--480, 2010.

\bibitem[Fern\'{a}ndez and Steel(1998)]{Fernandez}
C.~Fern\'{a}ndez and M.~F.~J. Steel.
\newblock On bayesian modeling of fat tails and skewness.
\newblock \emph{Journal of the American Statistical Association}, 93\penalty0
  (441):\penalty0 359--371, 1998.

\bibitem[Fr\"{u}hwirth-Schnatter and Lopes(2018)]{SchnatterLopes2018}
S.~Fr\"{u}hwirth-Schnatter and H.~F. Lopes.
\newblock Sparse bayesian factor analysis when the number of factors is
  unknown.
\newblock \emph{arXiv: 1804.04231}, 2018.

\bibitem[Geweke and Zhou(1996)]{GewekeZhou}
J.~Geweke and G.~Zhou.
\newblock Measuring the pricing error of the arbitrage pricing theory.
\newblock \emph{The Review of Financial Studies}, 9\penalty0 (2):\penalty0
  557--587, 1996.

\bibitem[Hansen(1994)]{Hansen}
B.~E. Hansen.
\newblock Autoregressive conditional density estimation.
\newblock \emph{International Economic Review}, 35\penalty0 (3):\penalty0
  705--730, 1994.

\bibitem[Harvey et~al.(2010)Harvey, Liechty, Liechty, and
  M\"{u}ller]{Harvey2010}
C.~R. Harvey, J.~C. Liechty, M.~W. Liechty, and P.~M\"{u}ller.
\newblock Portfolio selection with higher moments.
\newblock \emph{Quantitative Finance}, 10\penalty0 (5):\penalty0 469--485,
  2010.

\bibitem[Kon(1984)]{Kon}
S.~J. Kon.
\newblock Models of stock returns---a comparison.
\newblock \emph{The Journal of Finance}, 39\penalty0 (1):\penalty0 147--165,
  1984.

\bibitem[Li et~al.(2019)Li, Craig, and Bhadra]{Li2019}
Y.~Li, B.~A. Craig, and A.~Bhadra.
\newblock The graphical horseshoe estimator for inverse covariance matrices.
\newblock \emph{Journal of Computational and Graphical Statistics}, 28\penalty0
  (3):\penalty0 747--757, 2019.

\bibitem[Lopes and West(2004)]{LopesWest2004}
H.~F. Lopes and M.~West.
\newblock Bayesian model assessment in factor analysis.
\newblock \emph{Statistica Sinica}, 14\penalty0 (1):\penalty0 41--67, 2004.

\bibitem[Makalic and Schmidt(2016)]{MakalicSchmidt2016}
E.~Makalic and D.~F. Schmidt.
\newblock A simple sampler for the horseshoe estimator.
\newblock \emph{IEEE Signal Processing Letters}, 23\penalty0 (1):\penalty0
  179--182, 2016.

\bibitem[Markowitz(1952)]{Markowitz}
H.~Markowitz.
\newblock Portfolio selection.
\newblock \emph{Journal of Finance}, 7\penalty0 (1):\penalty0 77--91, 1952.

\bibitem[Markowitz and Usmen(1996)]{Markowitz96}
H.~Markowitz and N.~Usmen.
\newblock The likelihood of various stock market return distributions, part 2:
  Empirical results.
\newblock \emph{Journal of Risk and Uncertainty}, 13\penalty0 (3):\penalty0
  221--247, 1996.

\bibitem[Mills(1995)]{Mills}
T.C. Mills.
\newblock Modelling skewness and kurtosis in the london stock exchange ft-se
  index return distributions.
\newblock \emph{Journal of the Royal Statistical Society: Series D (The
  Statistician)}, 44\penalty0 (3):\penalty0 323--332, 1995.

\bibitem[Nakajima(2017)]{Nakajima2017}
J.~Nakajima.
\newblock Bayesian analysis of multivariate stochastic volatility with skew
  return distribution.
\newblock \emph{Econometric Reviews}, 36\penalty0 (5):\penalty0 546--562, 2017.

\bibitem[Nakajima and Omori(2012)]{NakajimaOmori}
J.~Nakajima and Y.~Omori.
\newblock Stochastic volatility model with leverage and asymmetrically
  heavy-tailed error using gh skew student's t-distribution.
\newblock \emph{Computational Statistics \& Data Analysis}, 56\penalty0
  (11):\penalty0 3690--3704, 2012.

\bibitem[Oya and Nakatsuma(2021)]{ONGLASSO}
S.~Oya and T.~Nakatsuma.
\newblock A positive-definiteness-assured block gibbs sampler for bayesian
  graphical models with shrinkage priors.
\newblock \emph{arXiv:2001.04657v2}, 2021.

\bibitem[Panagiotelis and Smith(2010)]{PS2010}
A.~Panagiotelis and M.~Smith.
\newblock Bayesian skew selection for multivariate models.
\newblock \emph{Computational Statistics \& Data Analysis}, 54\penalty0
  (7):\penalty0 1824--1839, 2010.

\bibitem[Peir\'{o}(1999)]{Peiro}
A.~Peir\'{o}.
\newblock Skewness in financial returns.
\newblock \emph{Journal of Banking \& Finance}, 23\penalty0 (6):\penalty0
  847--862, 1999.

\bibitem[Sahu et~al.(2003)Sahu, Dey, and Branco]{Sahu03}
S.~K. Sahu, D.~K. Dey, and M.D. Branco.
\newblock A new class of multivariate skew distributions with applications to
  bayesian regression models.
\newblock \emph{Canadian Journal of Statistics}, 31\penalty0 (2):\penalty0
  129--150, 2003.

\bibitem[Wang(2012)]{Wang}
H.~Wang.
\newblock Bayesian graphical lasso methods and efficient posterior computation.
\newblock \emph{Bayesian Analysis}, 7:\penalty0 867--886, 2012.

\bibitem[Watanabe(2001)]{WATANABE2001177}
T.~Watanabe.
\newblock On sampling the degree-of-freedom of student's-t disturbances.
\newblock \emph{Statistics \& Probability Letters}, 52\penalty0 (2):\penalty0
  177--181, 2001.

\bibitem[West(2003)]{West2003}
M.~West.
\newblock Bayesian factor regression models in the ``large p, small n''
  paradigm.
\newblock \emph{Bayesian Statistics}, 7:\penalty0 733--742, 2003.

\end{thebibliography}
\bibliographystyle{plainnat}

\section*{Appendix}
\label{appendix}
As we mentioned before, \cite{Harvey2010} developed the Gibbs sampling algorithm for the multivariate skew-normal distribution \eqref{sn.def1}, but it is straightforward to extend it to the multivariate skew-t distribution as \cite{Sahu03} showed. Since it is expressed as a scale mixture of multivariate skew-normal distributions, skew-t distributed $R_{t}$ is expressed as
\begin{equation}
    \label{st.def}
    \begin{split}
        & R_t = \mu + \Delta Z_t + \epsilon_t, \\
        & Z_t\sim\mathcal{N}^+\left(0, \frac1{\gamma_t}I_N\right),\quad \epsilon_t\sim\mathcal{N}\left(0,\left(\gamma_t\Omega\right)^{-1}\right),\quad \gamma_t\sim Ga\left(\frac{\varphi}2,\frac{\varphi}2\right), \\
        & Z_t\perp \epsilon_t\perp \gamma_t,
    \end{split}
\end{equation}
Given $\gamma_t$, the sampling algorithms for $\delta$, $\Omega$, $\mu$ and $Z_t$ in \eqref{st.def} are almost identical to the multivariate skew-normal case except that
\begin{description}
    \item[$\delta$:] redefine $\hat A_\delta$ and $\hat b_\delta$ in \eqref{fc.delta} as
\[
    \hat A_\delta = A_\delta + \sum_{t=1}^T\gamma_t W_t^{\intercal}\Omega W_t,\quad
    \hat b_\delta = A_\delta b_\delta + \sum_{t=1}^T\gamma_t W_t^{\intercal}\Omega\tilde R_t.
\]
    \item[$\Omega$:] redefine $S$ in \eqref{fc.omega} as
\[
    S = \sum_{t=1}^T\gamma_t(R_t-\mu-\Delta Z_t)(R_t-\mu-\Delta Z_t)^{\intercal}.
\]
    \item[$\mu$:] redefine $\hat A_\mu$ and $\hat b_\mu$ in \eqref{fc.mu} as
\[
    \hat A_\mu = A_\mu + \sum_{t=1}^T\gamma_t\Omega,\quad
    \hat b_\mu = A_\mu b_\mu + \sum_{t=1}^T\gamma_t\Omega(R_t - \Delta Z_t).
\]
    \item[$Z_t$:] redefine $\hat A_z$ and $\hat b_z$ in \eqref{fc.latent} as
\[
    \hat A_z = \gamma_t\left(I_N + \Delta^{\intercal}\Omega\Delta\right),\quad
    \hat b_z = \gamma_t\Delta^{\intercal}\Omega(R_t-\mu).
\] 
\end{description}
Finally, with the prior $\varphi\sim Ga(a_\varphi,b_\varphi)$, the full conditional posterior distribution of $\varphi$ is derived as
\begin{align}
    \label{fc.varphi}
    p(\varphi|\cdot) &\propto \prod_{t=1}^T\frac{\left(\frac{\varphi}2\right)^{\frac{\varphi}2}}{\Gamma\left(\frac{\varphi}2\right)}\gamma_t^{\frac{\varphi}2-1}\exp\left(-\frac{\varphi\gamma_t}2\right)\times \varphi^{a_\varphi-1}\exp(-b_\varphi\varphi) \nonumber\\
    &\propto \frac{\left(\frac{\varphi}2\right)^{\frac{\varphi T}2}}{\Gamma\left(\frac{\varphi}2\right)^T}\left(\prod_{t=1}^T\gamma_t\right)^{\frac{\varphi}2-1}\exp\left(-\frac{\varphi}2\sum_{t=1}^T\gamma_t\right)\times \varphi^{a_\varphi-1}\exp(-b_\varphi\varphi) \nonumber\\
    &\propto \exp\left[\left(\frac{\varphi T}2 + a_\varphi - 1\right)\log\varphi - T\log\Gamma\left(\frac{\varphi}2\right) - \hat b_\varphi\varphi\right],
\end{align}
where
\[
    \hat b_\varphi = b_\varphi + \frac{\log 2}2T + \frac12\sum_{t=1}^T\left(\gamma_t - \log\gamma_t\right).
  \]
Following \cite{WATANABE2001177}, we may apply a Metropolis-Hastings algorithm to draw $\varphi$ from \eqref{fc.varphi}. For this purpose, we consider the second-order Taylor approximation of
\[
    f(\varphi) = \left(\frac{\varphi T}2 + a_\varphi - 1\right)\log\varphi - T\log\Gamma\left(\frac{\varphi}2\right) - \hat b_\varphi\varphi,
\]
within the exponential function of \eqref{fc.varphi}, that is,
\[
    f(\varphi) \approx f(\varphi^*) + \nabla f(\varphi^*)(\varphi-\varphi^*) + \frac12\nabla^2 f(\varphi^*)(\varphi-\varphi^*),
\]
where
\begin{align*}
    \nabla f(\varphi) &= \frac{T}2\log\varphi + \frac{T}2 + \frac{a_\varphi - 1}{\varphi} - \frac{T}2\nabla\log\Gamma\left(\frac{\varphi}2\right) - \hat b_\varphi, \\
    \nabla^2 f(\varphi) &= \frac{T}2\left(\frac1{\varphi}-\frac12\nabla^2\log\Gamma\left(\frac{\varphi}2\right)\right) - \frac{a_\varphi - 1}{\varphi^2}.
\end{align*}
Note that $f$ is globally concave and has a unique mode. If we take the mode of $f$ as $\varphi^*$, we have $\nabla f(\varphi^*)=0$. Thus the pdf of the full conditional posterior distribution \eqref{fc.varphi} is approximated as
\[
    p(\varphi|\cdot) \approx \mathcal{K} \exp\left[\frac12\nabla^2 f(\varphi^*)(\varphi-\varphi^*)\right],
\]
Therefore we can use
\[
    \varphi \sim \mathcal{N}^+\left(\varphi^*,\ \left\{-\nabla^2 f(\varphi^*)\right\}^{-1}\right)
\]
as the proposal distribution of $\varphi$ in the Metropolis-Hastings algorithm.

\end{document}